\documentclass[iop, numberedappendix]{mn2e}

\usepackage[pdftex]{graphicx}

\usepackage{amsmath, amsthm, amssymb}
\let\bibhang\relax
\usepackage{natbib}
\usepackage{subfigure}
\usepackage{upgreek}
\usepackage{accents}
\usepackage{color}
\usepackage{hyperref}
\usepackage{dblfloatfix}
\usepackage[T1]{fontenc}
\usepackage{aecompl}
\usepackage{enumitem}

\hypersetup{
  colorlinks   = true, 
  urlcolor     = blue, 
  linkcolor    = blue, 
  citecolor   = blue 
}

\newcommand{\ares}{\textsc{ares}}

\newcommand{\angstrom}{\text{\normalfont\AA}}

\newcommand{\Mpch}{\text{Mpc} / h}
\newcommand{\Mpchinv}{h \ \text{Mpc}^{-1}}

\newcommand{\Rmfp}{R_{\text{mfp}}}

\newcommand{\mAB}{m_{\text{AB}}}

\newcommand{\zreion}{z_{\text{reion}}}

\newcommand{\tot}{\text{tot}}

\newcommand{\MARtot}{\dot{M}_{h,\tot}}
\newcommand{\MARacc}{\dot{M}_{h,\text{acc}}}
\newcommand{\MARmer}{\dot{M}_{h,\text{mer}}}
\newcommand{\MARunits}{M_{\odot} \ \text{yr}^{-1}}

\newcommand{\ham}{\texttt{ham}}

\newcommand{\univ}{\texttt{univ}}
\newcommand{\univinst}{\texttt{univ-inst}}
\newcommand{\univhist}{\texttt{univ-hist}}

\newcommand{\lsfrunits}{\text{erg} \ \text{s}^{-1} \ \text{Hz}^{-1} \ (M_{\odot} \ \text{yr}^{-1})^{-1}}

\newcommand{\alphalo}{\alpha_{\mathrm{lo}}}
\newcommand{\alphahi}{\alpha_{\mathrm{hi}}}
\newcommand{\alphaloD}{\alpha_{d,\mathrm{lo}}}
\newcommand{\alphahiD}{\alpha_{d,\mathrm{hi}}}

\newcommand{\fdtmr}{f_{\text{dtmr}}}

\newcommand{\Mmin}{M_{\min}}
\newcommand{\Mres}{M_{\text{res}}}

\newcommand{\fstar}{f_{\ast}}

\newcommand{\fcoll}{f_{\text{coll}}}

\newcommand{\Nion}{N_{\text{ion}}}
\newcommand{\fesc}{f_{\text{esc}}}

\newcommand{\Msun}{M_{\odot}}

\newcommand{\AUV}{A_{\text{UV}}}
\newcommand{\MUV}{M_{\text{UV}}}

\title[Galaxy histories during reionization]{The importance of galaxy formation histories in models of reionization}
\author[Mirocha, La Plante, \& Liu]{
Jordan Mirocha,$^{1}$\textsuperscript{\thanks{jordan.mirocha@mcgill.ca}}\textsuperscript{\thanks{CITA National Fellow}}
Paul La Plante, $^{2}$\textsuperscript{\thanks{BCCP Fellow}}  and
Adrian Liu$^{1}$ \\
$^{1}$McGill University Department of Physics \& McGill Space Institute, 3600 Rue University, Montr\'eal, QC, H3A 2T8 \\
$^{2}$Department of Astronomy and Radio Astronomy Laboratory, University of California Berkeley, Berkeley, CA 94720, USA \\
}

\begin{document}

\pagerange{\pageref{firstpage}--\pageref{lastpage}} \pubyear{2016}
\maketitle

\begin{abstract}
Upcoming galaxy surveys and 21-cm experiments targeting high redshifts $z\gtrsim 6$ are highly complementary probes of galaxy formation and reionization. However, in order to expedite the large volume simulations relevant for 21-cm observations, many models of galaxies within reionization codes are entirely subgrid and/or rely on halo abundances only. In this work, we explore the extent to which resolving and modeling individual galaxy formation histories affects predictions both for the galaxy populations detectable by upcoming surveys and the signatures of reionization accessible to upcoming 21-cm experiments. We find that a common approach, in which galaxy luminosity is assumed to be a function of halo mass only, is biased with respect to models in which galaxy properties are evolved through time via semi-analytic modeling and thus reflective of the diversity of assembly histories that naturally arise in $N$-body simulations. The diversity of galaxy formation histories also results in scenarios in which the brightest galaxies do \textit{not} always reside in the centers of large ionized regions, as there are often relatively low-mass halos undergoing dramatic, but short-term, growth. This has clear implications for attempts to detect or validate the 21-cm background via cross correlation. Finally, we show that a hybrid approach -- in which only halos hosting galaxies bright enough to be detected in surveys are modeled in detail, with the rest modeled as an unresolved field of halos with abundance related to large-scale overdensity -- is a viable way to generate large-volume `simulations` well-suited to wide-area surveys and current-generation 21-cm experiments targeting relatively large $k \lesssim 1 \ \Mpchinv$ scales.
\end{abstract}
\begin{keywords}
galaxies: high-redshift -- intergalactic medium -- galaxies: luminosity function, mass function -- dark ages, reionization, first stars -- diffuse radiation.
\end{keywords}

\section{Introduction} \label{sec:intro}
Enormous progress in modeling the Epoch of Reionization (EoR) has been made in the last two decades, including new insights drawn from analytical models \citep{Furlanetto2004}, computational efficiency boosts via so-called ``semi-numerical'' algorithms \citep{Mesinger2007,Mesinger2011}, and the growing detail of predictions made possible by full-fledged radiative transfer simulations in cosmological volumes \citep[e.g.,][]{Sokasian2001,Iliev2006,McQuinn2007,trac_cen2007}. Our understanding of the likeliest sources of reionization -- star-forming galaxies -- has also improved tremendously over the same timeframe, both empirically, largely via programs on the \textit{Hubble Space Telescope} \citep[HST; e.g.,][]{Windhorst2011,Koekemoer2011,Grogin2011,Bouwens2011,Illingworth2013}, and theoretically, via semi-analytic models \citep[see reviews by, e.g.,][]{Benson2010,Somerville2015} and \textit{ab initio} simulations of galaxy formation \citep[e.g.,][]{OShea2015,Schaye2015,Hopkins2014,Lovell2020}.

Of course, the processes of galaxy formation and reionization are intimately linked, as galaxies drive reionization while reionization itself can feed back on galaxy formation \citep{Shapiro1994,Gnedin2000,Bullock2000,Noh2014}. Treating this link explicitly is exceedingly challenging, as it requires capturing the ``full'' dynamic range of the reionization problem, ideally resolving the collapse of individual giant molecular clouds within galaxies ($\lesssim 10$ pc) within a suitably-large cosmological volume ($\gtrsim 100$ Mpc) to enable statistically-robust predictions \citep{Iliev2014}, all while including the relevant physics, e.g., primordial chemistry, hydrodynamics, and radiative transfer of ionizing photons. While impressive progress has been made in all of these areas \citep[e.g.,][]{Wise2011,Ocvirk2016,Smith2017,Trac2015,Gnedin2016}, finite computational resources will always require trade-offs between physics, spatial resolution, and simulation volume. In order to jointly model galaxies and reionization, at least approximately, volume is generally the most important consideration given that reionization drives 21-cm features on large $\gtrsim 10$ Mpc scales \citep{Furlanetto2004}, the same scales that are most readily accessible with current facilities. Galaxies must then be modeled via subgrid recipes.

In recent years, many models have employed abundance matching in order to quickly generate realistic high-$z$ galaxy populations \citep[e.g.,][]{Mason2015,Sun2016,Mashian2016}. Abundance matching reverse engineers the mass-to-light ratio of galaxies assuming that there is a 1:1 galaxy:halo correspondence, in which case one can associate galaxies of a particular luminosity by matching their abundance to model halos of a particular mass. This is an economical way to build what we will hereafter refer to as ``constrained'' models of reionization, because galaxies populate halos in a way that is constrained to agree with current observations by construction. Abundance matching allows one to, e.g., divert the majority of computational resources to performing radiative transfer in a cosmological volume, as one can sacrifice resolution on galaxy scales without sacrificing the ability to generate realistic galaxy populations \citep[e.g.,][]{McQuinn2007,Trac2015}. This is possible even in large-volume semi-numeric models in which the halos hosting high-$z$ galaxies are not resolved; in this case, the density field in suitably-large voxels can be used to predict the halo abundance \citep{McQuinn2007,Mesinger2007}.

While abundance matching is likely the most efficient way to populate halos with galaxies, it is of course not the only way or the most realistic way. Semi-analytic models take a more physically-motivated approach, and can generate galaxies whose properties reflect the diversity of dark matter halo assembly histories. As we will find in \S\ref{sec:results}, the neglect of halo histories acts as a systematic uncertainty in simpler models. There is, of course, a cost associated with avoiding this systematic through more detailed modeling of halo growth: the mass resolution required to capture the histories of all halos important for reionization cannot generally be achieved in a simulation volume large enough to generate mock observables. Most semi-numerical models of reionization circumvent this problem by modeling the halo population in aggregate only \citep[though see, e.g.,][]{Mutch2016,Mondal2017,Hutter2020}, relating the abundance of halos directly to the linearly-evolved density field \citep{Mesinger2007,Choudhury2009,Mesinger2011,Santos2010,Fialkov2013}. This approach trades resolution for efficiency, as voxels must be $\sim 1-4$ Mpc on a side in order to capture a representative halo population. Given that upcoming 21-cm observations target large, $\gtrsim 1$ Mpc scales, this trade is well justified.

In this work, we explore the effects of including galaxy histories in reionization models, and investigate the potential for a new kind of ``hybrid'' reionization model in which only the massive halos -- most likely to host the bright galaxies accessible to galaxy surveys -- are modeled in detail, while low-mass halos are modeled in aggregate. Such a model could allow one to simultaneously model directly-observable galaxies in detail without missing the ionizing photons of fainter source populations. Computationally, this could dramatically reduce the number of detailed galaxy histories being modeled, and thus make Markov Chain Monte-Carlo (MCMC) explorations of parameter space feasible. In addition, such an approach could be a powerful tool for galaxy--21-cm cross correlation modeling and interpreting constraints on individual galaxies in sky regions with overlapping 21-cm coverage. Finally, it will serve as an important test of different source modeling prescriptions, since the success of the model relies on our ability to accurately meld together two common approaches: those which treat the formation history of halos and those which do not.

We outline the different methods for modeling high-$z$ galaxies operating within our hybrid model in \ref{sec:galaxies}. In \S\ref{sec:results}, we compare these different galaxy modeling prescriptions and test the accuracy of their predictions for reionization, including the mean history, Thomson scattering optical depth, and 21-cm power spectrum. In \S\ref{sec:conclusions}, we discuss the implications of these results for reionization modeling in general and summarize our key conclusions.

We adopt AB magnitudes throughout \citep{Oke1983}, i.e.,
\begin{equation}
    M_{\lambda} = -2.5 \log_{10} \left(\frac{f_{\lambda}}{3631 \ \mathrm{Jy}} \right)
\end{equation}
and adopt the following cosmology: $\Omega_m = 0.3156$, $\Omega_b = 0.0491$, $h = 0.6726$, and $\sigma_8=0.8159$, very similar to the recent \citet{Planck2018} constraints.

\section{Galaxy Modeling} \label{sec:galaxies}
Because much of the diversity in model galaxy populations is driven by diversity in the dark matter halo population, we first describe our $N$-body simulations and the halo populations we extract from them in \S\ref{sec:halos}. Then, we introduce the galaxy modeling schemes we employ for $N$-body halos and simpler models common in abundance matching schemes \S\ref{sec:galaxies}. Finally, we describe our basic semi-numerical approach to reionization in \S\ref{sec:seminumerics}, before moving onto our results in \S\ref{sec:results}.

\subsection{The Halo Population} \label{sec:halos}
We employ two techniques for modeling dark matter halos:
\begin{itemize}
  \item A fully numerical approach, in which halo merger trees are constructed from $N$-body simulations. Our main results are generated from a $(80 \ h^{-1} \mathrm{Mpc})^3$, $2048^3$ particle $N$-body simulation, though we use smaller boxes with equal (or greater) mass resolution -- $(40 \ h^{-1} \mathrm{Mpc})^3$ at $1024^3$ and $2048^3$ resolution, for testing and convergence studies (see Appendix \ref{sec:convergence}). Each simulation is run to $z \simeq 5$.
  \item A (nearly) analytic approach, in which we adopt the \citet{Tinker2010} mass function and derive halo growth histories via abundance matching the HMF with itself at subsequent timesteps. This technique was first described in \citet{Furlanetto2017} -- it effectively assumes that halos are neither created or destroyed, and so the abundance evolution of halos encodes their mass evolution. Though simple, it provides good agreement with the average halo mass accretion rates (MARs) derived from simulations \citep[e.g.,][]{McBride2009,Trac2015}, which makes sense, at least to first order, as the growth of high-$z$ halos in simulations is dominated by inflow rather than mergers \citep[e.g.,][]{Goerdt2015}.
\end{itemize}
The second approach is the default in the \textsc{ares} code\footnote{\url{https://ares.readthedocs.io/en/latest/}}, which we use for all semi-analytic modeling. It has the advantage of being computationally efficient, given that one need only model galaxies in $\sim 10^3$ mass bins (since intra-bin diversity is effectively neglected), whereas our $N$-body simulations have $\sim 10^6$-$10^7$ halos, depending on redshift. As a result, we use the best-fitting galaxy parameters from MCMCs run with \ares\ as initial guesses for models based on $N$-body simulations (see \S\ref{sec:galaxies}).

A few more words about the $N$-body simulations and merger trees are warranted before we proceed with a comparison of each method. The $N$-body simulations employed in this work use a P$^3$M algorithm described
in \citet{Trac2015} for tracking particle positions and velocities. It also
constructs halo catalogs on-the-fly, which are generated every 20 Myr in cosmic
time. These halo catalogs are initially found using a friend-of-friends \citep[FOF;][]{Davis1985}
algorithm with 20 particles as the threshold for generating catalogs. These FoF
halos are then converted to spherical overdensity (SO) halos, which form the
basis of the halos for the halo catalog. The fundamental particle mass
resolution $m_p$ for the simulation resolutions described above is roughly
$m_p = 5.22 \times 10^6$ $h^{-1}M_\odot$. With a minimum-mass halo consisting of
20 particles, the simulation resolves halos with a mass of
$M_h \geq 1.04 \times 10^8$ $h^{-1}M_\odot$. Halo membership of particles is
also tracked, which allows for the construction of halo merger trees, discussed
more in Sec.~\ref{sec:rhs}.

\subsubsection{Resolved Halos} \label{sec:rhs}
\begin{figure}
\begin{center}
\includegraphics[width=0.49\textwidth]{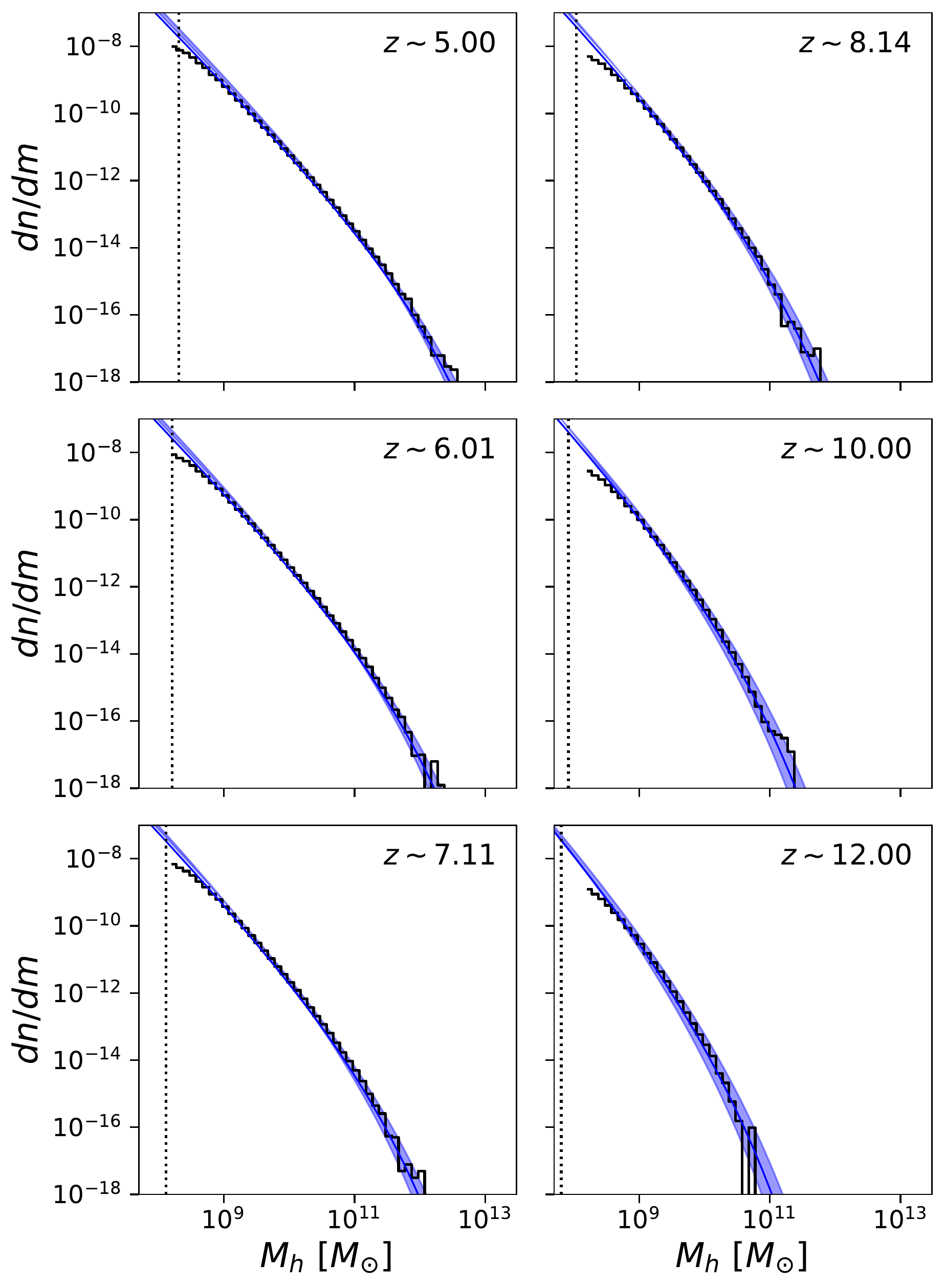}
\caption{{\bf Halo mass function of simulated halos agrees well with the \citetalias{Tinker2010} form, and is complete (or nearly complete) for all $M_h \gtrsim 10^8 \ \Msun$.} Shaded contours bound the range between \citetalias{PressSchechter1974} (lower edge) and \citetalias{ShethMoTormen2001} (upper edge) mass functions, while the solid blue line shows the \citetalias{Tinker2010} mass function. Solid black histograms are simulation results. Vertical dotted line indicates atomic cooling threshold, i.e., halo mass corresponding to a virial temperature of $10^4$ K, at each redshift.}
\label{fig:hmf}
\end{center}
\end{figure}

In Figure \ref{fig:hmf}, we compare the halo mass function (HMF) derived from our simulations to the \citet{Tinker2010} (hereafter T10), \citet{ShethMoTormen2001} (hereafter SMT), and \citet{PressSchechter1974} (hereafter PS) fitting functions. We find good agreement at all redshifts, with some incompleteness visible in halos just above the atomic cooling threshold, $M_h \lesssim 10^{8.5} \ \Msun$, and fluctuations in massive halo counts due to the finite size of the simulation volume (i.e., cosmic variance).

\defcitealias{PressSchechter1974}{PS}
\defcitealias{ShethMoTormen2001}{SMT}
\defcitealias{Tinker2010}{T10}

Beause our models use the histories of dark matter halos, not just their masses, we now describe our simple approach to building merger trees. There is a substantial literature on this topic alone, for both analytic and numerical methods \citep[e.g.,][]{Lacey1994,Somerville1999,Behroozi2013Trees,Poole2017}, so we summarize only the pertinent details for this work. The FoF halo finding procedure yields a catalog of halos at each snapshot, as well as a list of particle IDs for all progenitor halos in the previous snapshot. We perform a recursive search of each branch, i.e., starting from the final snapshot at $z\sim 5$, for each halo we ascend through previous snapshots and assemble a list of progenitor halos (and their progenitors, and so on). After this first pass, we have a preliminary set of halo histories, as well as a mapping between progenitors and descendants at each timestep. There are many potential ``pathologies'' present in this first-pass merger tree, as have been discussed in great detail elsewhere \citep[see, e.g.,][]{Poole2017}.

After assembling an initial merger tree, we descend back through the tree along each branch, from high redshift to low, and use the inferred mass growth rate, $\MARtot$, to identify potential artifacts. We subdivide the total growth into growth by mergers and accretion, denoted $\MARmer$ and  $\MARacc$, respectively. The growth via accretion is inferred simply as the difference between total growth between a timestep and that attributable to mergers, i.e., $\MARacc = \MARtot - \MARmer$. We assume that \textit{baryonic} mass inflow always reflects the cosmic baryon fraction, i.e., the baryonic MAR $\dot{M}_b = f_b \MARacc$, where $f_b \equiv \Omega_b / \Omega_m$.

Now, due to a variety of potential halo-halo interactions and/or numerical artifacts near the resolution threshold, halo growth rates can become negative. This will of course wreak havoc on any MAR-based SFR recipe like ours if left unattended. In many cases, the origins of such excursions are difficult to pinpoint, as we do not attempt to track sub-halos using, e.g., \textsc{subfind} \citep{Springel2001}. We address such scenarios as follows. For small, $-0.1 \leq \Delta \log_{10}(\MARtot/[\MARunits]) < 0$ changes to a halo's mass, we simply null the MAR inferred for that timestep and update the mass accordingly. Larger deviations likely indicate ``dropped'' or ``bridged'' halos \citep{Poole2017}, i.e., halos that disappear for a timestep or two, either because they are near the resolution limit or are incorporated into a larger halo temporarily. In either case, when such halos re-emerge, we will derive a large, unphysical MARs from the initial merger tree. To combat this, we simply null the MAR over the entire interval in which these halos are missing from the tree, and replace the halo's mass with the average of neighboring grid points.

\begin{figure*}
\begin{center}
\includegraphics[width=0.98\textwidth]{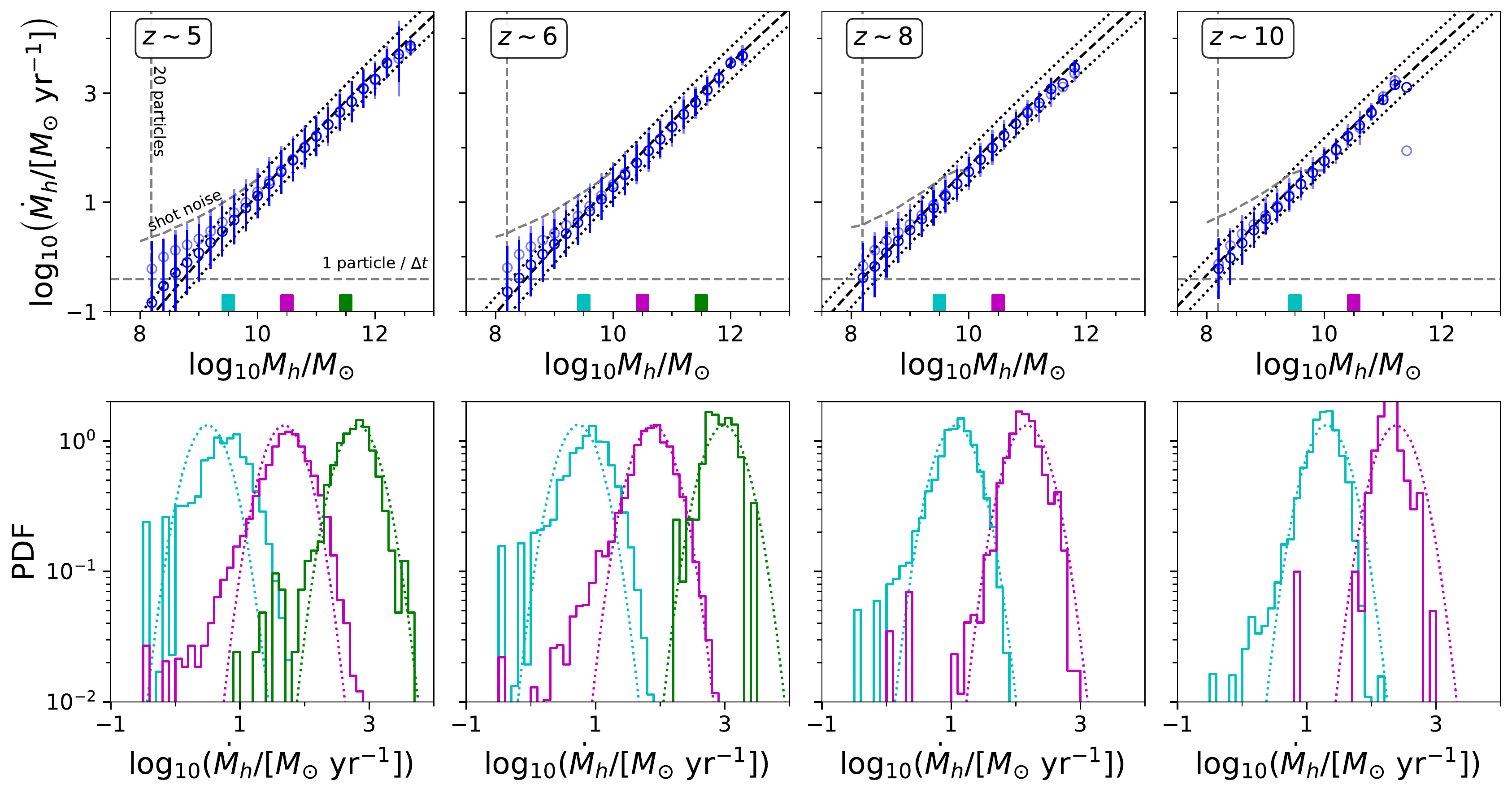}
\caption{{\bf Mass accretion rates of simulated halos from $z \sim 5$ to $z\sim 10$ (left to right) agree reasonably well with \citet{Furlanetto2017} approach} (dashed black), supporting idea that growth is largely through pristine inflow at high redshift. Top panels show the MAR averaged in a series of halo mass bins (points with error-bars), while bottom panels show the full distribution of the MAR in three 0.2 dex mass bins centered on $\log_{10} M_h = 9.5, 10.5$, and 11.5 (cyan, magenta, green; indicated along $x$-axis in top row). The $1-\sigma$ log-normal scatter is $\sim 0.3$ dex, which we show as dotted lines in each panel to guide the eye, though the tails of the distribution for simulated halos is generally broader. Several points of reference are indicated via dashed lines in the top row, including the halo mass corresponding to 20 particles (vertical line), the accretion rate resolution (1 particle per timestep; horizontal line), as well as an estimate of shot noise in the accretion rate (see text for details). Smoothing of the MAR over 100 Myr timescales (opaque blue) does reduce the ``raw'' MAR (semi-transparent blue) at the low-mass end, as it accounts for timesteps in which the accretion rate is zero.}
\label{fig:hmar}
\end{center}
\end{figure*}

In Figure \ref{fig:hmar}, we compare the halo MAR\footnote{Note that the halo mass accretion rate is not necessarily equivalent to the \textit{galaxy} mass accretion rate, as a slew of complicated feedback processes can prevent baryons from reaching the interstellar medium. However, the halo accretion rate is a reasonable proxy for the galaxy accretion rate, and is likely an increasingly accurate proxy at high-$z$ \citep{vandeVoort2011}.} inferred from our simulations to that predicted by a model in which halos are assumed to grow at fixed number density \citep[as in][]{Furlanetto2017}. Along the top panel, we see that agreement is good in the mean, at least in halos with $M_h \gtrsim 10^9 M_{\odot}$, which are resolved with more than $\sim 100$ DM particles. The simulated halo population exhibits log-normal scatter in the MAR (at fixed $M_h$) at the level of $\sim 0.3$ dex, which we indicate with dotted lines in the top panel (compare to error bars) and bottom panel (compared to MAR PDF).

The horizontal dashed lines in the top row of Figure \ref{fig:hmar} indicates our accretion rate resolution, i.e., the accretion rate of a halo that changes mass by one DM particle over a single timestep (20 Myr). The diagonal dashed lines indicate an estimate of shot noise; we compute the probability of accreting $N$ particles in a given timestep $\Delta t$, assuming the mean MAR predicted by the \citet{Furlanetto2017} model. The lines are the upper boundary of this Poissonian probability distribution function in each halo mass bin, i.e., all of the Poissonian probability lies below this curve (defined by $\dot{M}_h = N_{\mathrm{max}} M_{\text{DM}} / \Delta t$). The flattening trend of $\dot{M}_h$ at low-mass is not inconsistent with this shot noise. To mitigate this effect, we smooth the MAR in all halos $M_h \leq 5 \times 10^9 M_{\odot}$ with a top-hat kernal 100 Myr wide (5 timesteps). This reduces the positive bias in the MAR incurred at low-mass (semi-transparent vs. opaque points), but does not completely reconcile the MAR with the analytic approach of \citet{Furlanetto2017}. See Appendix \ref{sec:convergence} for further justification of this approach. We only include halos resolved with $\geq 20$ particles in all that follows (indicated by dashed vertical lines in each panel of Figure \ref{fig:hmar}).

\begin{figure}
\begin{center}
\includegraphics[width=0.49\textwidth]{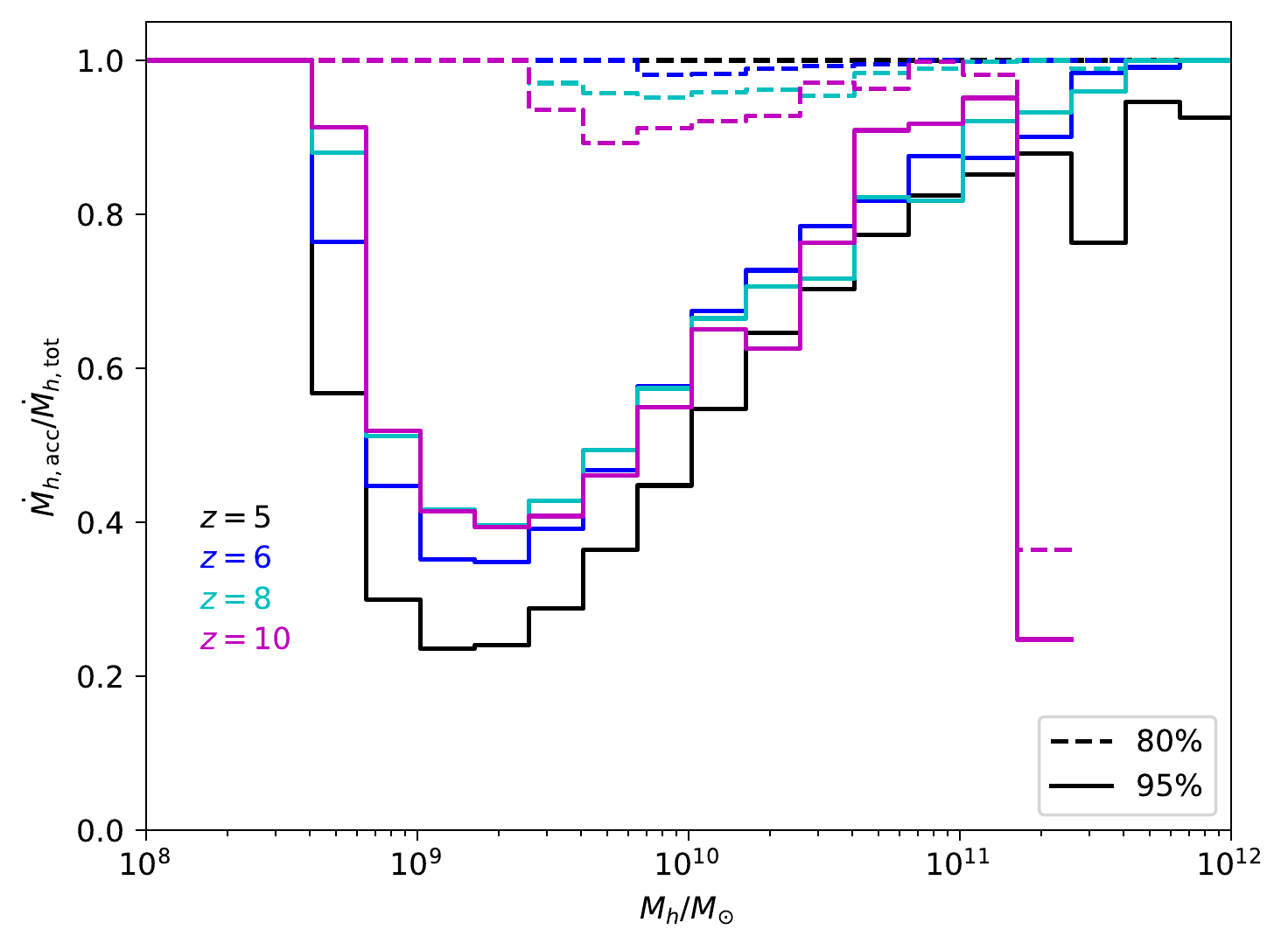}
\caption{{\bf Fraction of halo growth due to accretion.} In each halo mass bin, we show two contours: a solid line, above which 95\% of halos reside, and a dashed line, above which 80\% of halos reside. Colors indicate different redshifts between $z=5$ and $z=10$.}
\label{fig:mergers}
\end{center}
\end{figure}

Though the close agreement between simulated accretion rates and analytic predictions suggests that the effect of mergers is small, in agreement with other studies \citep[see, e.g.,][]{vandeVoort2011,Goerdt2015}, it is not zero. In Figure \ref{fig:mergers}, we show the fraction of halo growth that occurs via accretion, $f_{\mathrm{acc}}$, in a series of 0.2 dex halo mass bins at $z=5,6,8$ and 10. At all redshifts and masses, most growth is via accretion: 80\% of halos show $f_{\mathrm{acc}}$ values in excess of $\sim 0.9$, at all masses (dashed contours). Some halos do show evidence of recent mergers, though in 95\% of all cases, $\gtrsim 60$\% of the growth is still via accretion from the IGM (solid curves). Values equal to unity at $M_h \lesssim 10^{8.5} \ M_{\odot}$ in Figure \ref{fig:mergers} are a resolution effect -- halos in this range are too small to have merged with a resolved halo, and as a result, any mass growth must be due to inflow. The key conclusion from Figure \ref{fig:mergers} is that $\lesssim 20$\% of halos at any given snapshot have accrued more than $\sim 10$\% of their current mass from mergers. This will result in small under-estimate of galaxy luminosities in models neglecting mergers, since the luminosity of recently-merged progenitors is unaccounted for.

\begin{figure*}
\begin{center}
\includegraphics[width=0.98\textwidth]{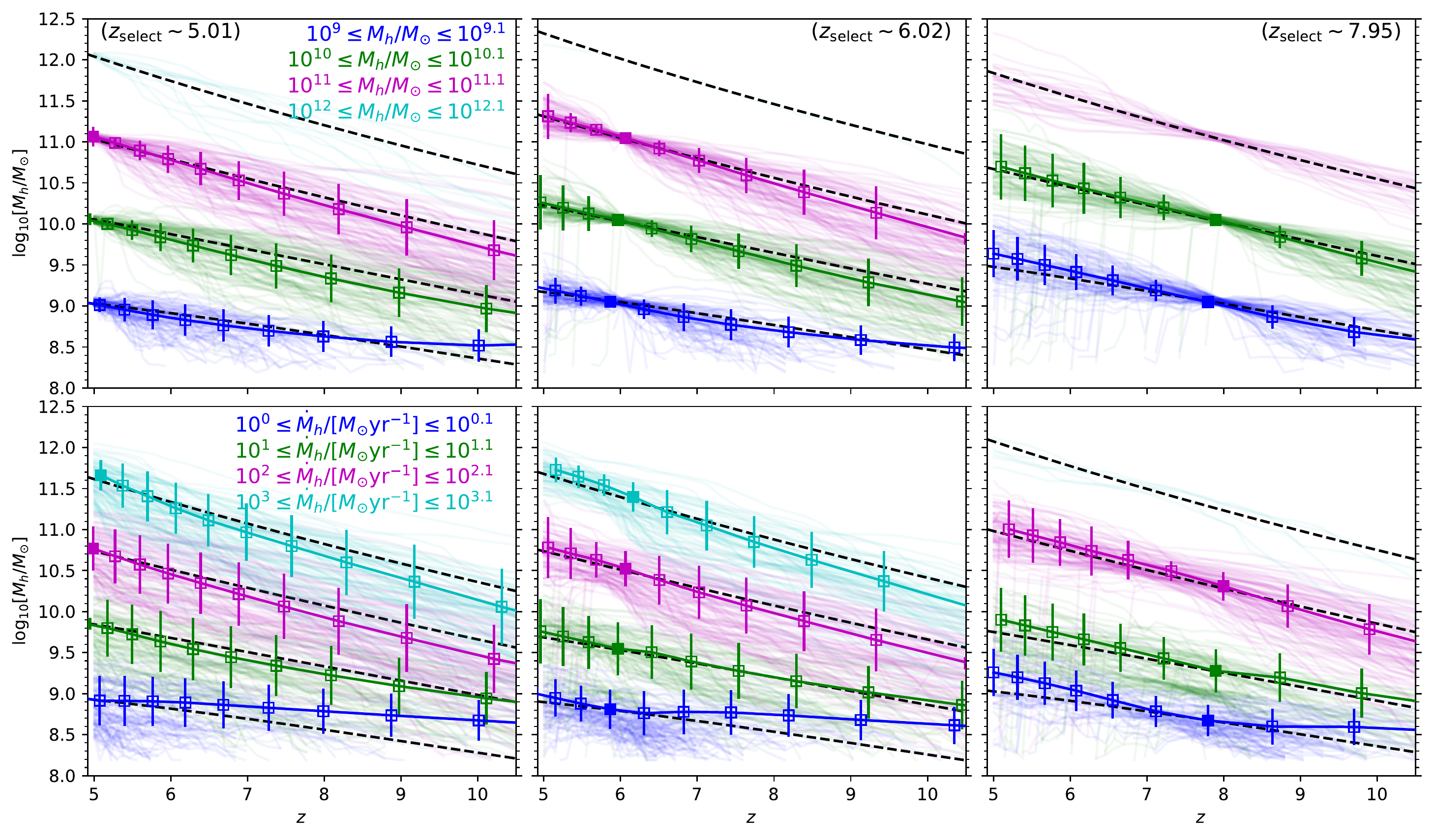}
\caption{{\bf Growth histories of simulated halos are generally steeper than the those of idealized halos evolving at fixed abundance, and exhibit more scatter at fixed MAR than fixed $M_h$.} \textit{Top:} Semi-transparent lines show 50 example histories for halos that have masses within a factor of 0.1 dex at $z \sim 5,6,$ and 8 (left, center, right), while points with error-bars show the mean halo mass at each snapshot for the selected population of objects (only shown for bins with more than 50 halos). Dashed lines show predicted growth histories from the \citet{Furlanetto2017} model, which neglects mergers and scatter in mass accretion rates. \textit{Bottom:} Analogous to top row, except halos are selected to have the same MAR (within 0.1 dex) at $z \sim 5,6,$ and 8. See \S\ref{sec:rhs} for details about the halo catalogs and accretion rates.}
\label{fig:hhist}
\end{center}
\end{figure*}

Though the halo MAR employed in simple models is in good agreement with the \textit{mean} growth rate of halos in simulations, the `trajectories' of individual halos differ in detail. As we will see shortly, this has an important impact on model calibration and reionization predictions. We explore this effect in Figure \ref{fig:hhist}, which plots the growth of `cohorts' of halos that resided in the same $0.1$ dex mass bin at some redshift, $z_{\text{select}} \sim 4$, 6, and 8 (left, center, right; top row). Clearly, halos of the same mass at one snapshot `diffuse' over time, and may not end up in the bin predicted by continuous growth at the mean MAR. This is a known phenomenon \citep[see, e.g.,][]{Behroozi2013progenitors,Torrey2017}, though not widely investigated at high-$z$ where abundance-based models for galaxies have become very prevalent. We will quantify the impact of this effect on our galaxy modeling in \S\ref{sec:results}. The bottom row of Figure \ref{fig:hhist} is analogous, but shows halos selected at fixed MAR, rather than $M_h$. Clearly, the scatter in $M_h$ and formation history at fixed MAR is considerable; this is very relevant for MAR-driven SFR models like ours, since galaxy luminosity is closely linked to MAR. In the lowest MAR selection bin (blue points; bottom row), limitations caused by finite resolution are prominent.

\subsubsection{Unresolved \& under-resolved halos} \label{sec:urhs}
Motivated by the potential importance of low-mass halos in reionization, we also explore the option of a hybrid approach in which halos above the resolution limit are modeled ``in full,'' while halos below this limit are modeled using approximate techniques. The resolution limit in this scheme is itself negotiable, as one may distrust the growth histories of halos resolved with $\lesssim 100$ particles, even though the halos masses are reasonably well-resolved (see Figure \ref{fig:hmar}). We will explore the effects of halos in the unresolved and marginally-resolved $M_h \lesssim 10^{10} M_{\odot}$ regime on reionization in \S\ref{sec:results}.

To begin \citep[see also, e.g., \S3 in][]{McQuinn2007}, we compute the matter overdensity $\delta$ on a coarse grid
with 1, 1.6, and $2 \ h^{-1}$Mpc resolution. This overdensity is converted to a total mass
$M_c$ by multiplying by the volume of the cell and the mean cosmic mass density
at $z = 0$. This Eulerian-space overdensity $\delta$ is also converted to the
Lagrangian-space overdensity $\delta_0$ using the fitting formula of
\citet{mo_white1996}:
\begin{equation}
  \delta_0(\delta, z) = \frac{1}{D(z)} \left[1.68647 - \frac{1.35}{\delta^{2/3}} - \frac{1.12431}{\delta^{1/2}} + \frac{0.78785}{\delta^{0.58661}}\right],
\end{equation}
where $D(z)$ is the growth factor at redshift $z$. Given the quantities
$\delta_0$ and $M_c$ for a particular voxel, we can compute the conditional mass
function using the \citetalias{PressSchechter1974}  mass function for
the expected number of halos $n_h$ of mass $M_h$:
\begin{align}
n_h(M_h, \delta_0, M_c, z) & = \sqrt{\frac{2}{\pi}} \frac{\bar{\rho}}{M_h^2} \left| \frac{d \log \sigma}{d \log M_h} \right| \frac{\delta_c(z) - \delta_0}{\sqrt{\sigma^2(M_h) - \sigma^2(M_c)}} \nonumber \\
& \times \exp\left(\frac{\left(\delta_c(z) - \delta_0\right)^2}{2\left(\sigma^2(M_h) - \sigma^2(M_c)\right)}\right),
\label{eqn:nh_ps}
\end{align}
where $\sigma^2(M_h)$ is the linear theory variance corresponding to a mass
$m$.

When applying this scheme to our $N$-body simulation outputs, we first compute
Equation~(\ref{eqn:nh_ps}) for all voxels in the volume. We compute the expected
number for several values of $M_h$, corresponding to halos of different masses
of interest. We take care to appropriately handle cases where $M_h < M_C$, so as
to avoid numerically nonsensical results.

%
\begin{figure*}
\begin{center}
\includegraphics[width=0.98\textwidth]{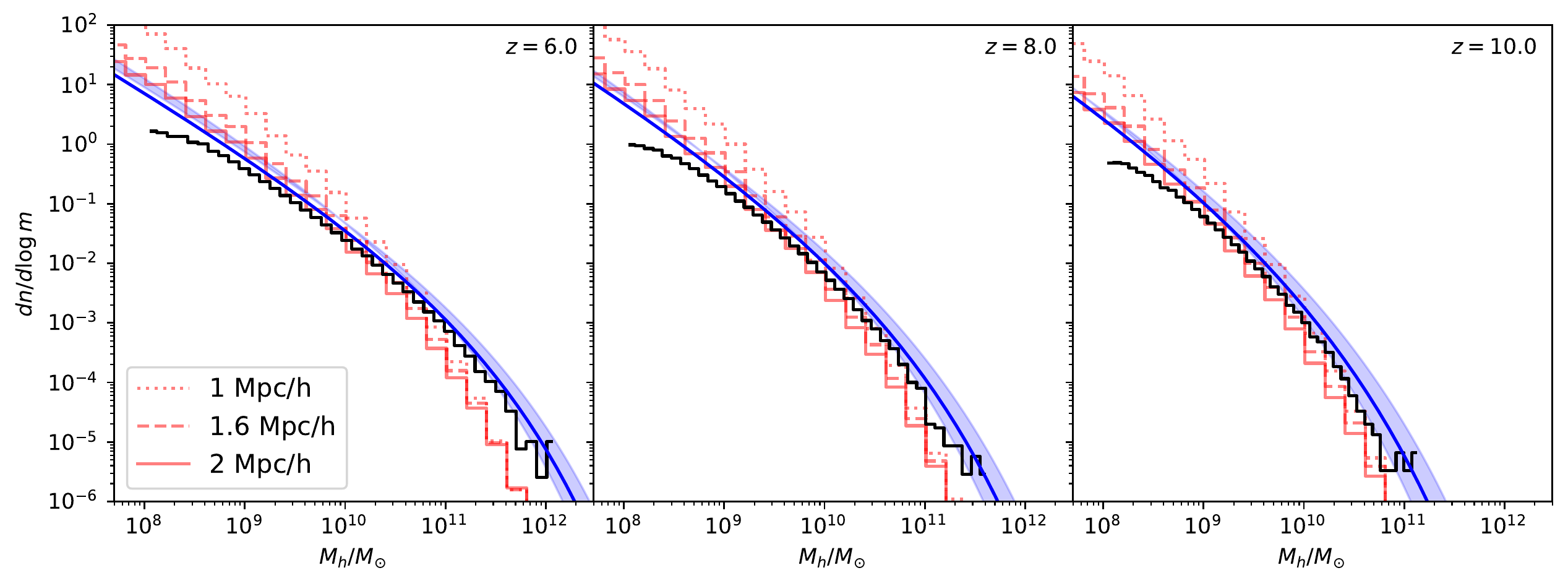}
\caption{{\bf Comparison of volume-averaged HMF computed from resolved halos (black) and from a model for unresolved halos (red).} As in Figure \ref{fig:hmf}, blue lines indicate the \citetalias{Tinker2010} (solid) and \citetalias{ShethMoTormen2001}-\citetalias{PressSchechter1974} mass functions (shaded). Dotted, dashed, and solid red curves apply Eq. \ref{eqn:nh_ps} on increasingly coarse grids, as indicated in the legend.}
\label{fig:urhs}
\end{center}
\end{figure*}

The results of this procedure are shown in Figure \ref{fig:urhs}. We see, as in Figure \ref{fig:hmf}, resolved halos in black compared to the named HMFs in blue, with the additional models for unresolved halos at $M_h \leq 10^{10} \Msun$ in red. We show three different cases adopting different grid resolutions of voxel size 1, 1.6, and 2 $h^{-1} \mathrm{Mpc}$, corresponding to $80^3$, $50^3$, and $40^3$ grids. Given the convergence of the HMF at $M_h \leq 10^{10} \ \Msun$ with the coarsest grid shown, we adopt $40^3$ grids for all subsequent models employing unresolved halos. Note that we make no effort to modify Eq. \ref{eqn:nh_ps} to reflect, e.g., the \citetalias{Tinker2010} form of the mass function, which agrees more closely with our simulated HMF. In addition, we do not attempt to synthesize merger histories for these unresolved halos, a point which we will revisit in \S\ref{sec:URGs}.

\subsection{Galaxy Formation} \label{sec:galaxies}
With a mock halo population in hand, whether constructed via simple analytic arguments or $N$-body simulations, we can proceed to generate mock galaxy populations. In each of our models, the star formation rate (SFR) is linked to pristine gas inflow, though the details of how the SFR is determined by the MAR differs slightly. In each of the following subsections, we provide a brief description of each model in turn, starting first with halo abundance matching (\ham), before moving on to more physically-motivated scenarios (\univ), all of which are implemented within the \textsc{ares} code \citep[see,e.g.,][]{Mirocha2017,Mirocha2020a}, and summarized briefly in Table \ref{tab:models}.

\begin{table*}
\begin{tabular}{ | l | l | l | }
\hline
model & dust & notes \\
\hline
\ham & IRX-$\beta$ \citep{Meurer1999} \& $\MUV$-$\beta$ \citep{Bouwens2014} & $L \propto \mathrm{SFR}$, halo abundance match to $z\sim 5$ UVLFs \\
\univinst & IRX-$\beta$ \citep{Meurer1999} \& $\MUV$-$\beta$ \citep{Bouwens2014} & $L \propto \mathrm{SFR}$, fit to UVLFs at $4 \lesssim z \lesssim 8$ \\
\univhist & forward model jointly fit UVLFs and $\MUV$-$\beta$ \citep{Mirocha2020a} & full spectral synthesis \\
\end{tabular}
\caption{{\bf Summary of source models used in this work.} Each model is calibrated to a subset of $4 \lesssim z \lesssim 8$ UVLFs from \citet{Bouwens2015} and $4 \lesssim z \lesssim 6$ $\MUV$-$\beta$ relations from \citet{Bouwens2014} using \ares. Each adopts a redshift-independent double power-law model for $f_{\ast}$ (Eq. \ref{eq:sfe_dpl}), but differs in the treatment of dust and whether the luminosity of galaxies is linked only to the current SFR or is synthesized over the entire star formation history.}
\label{tab:models}
\end{table*}

\subsubsection{Halo abundance Matching (\ham)} \label{sec:ham}
Our \ham\ model is very similar to others employed in the recent literature \citep[e.g.,][]{Trac2015,Sun2016}. We assume the rest-UV luminosity of galaxies is linearly related to the SFR, $\dot{M}_{\ast}$, i.e.,
\begin{equation}
	L_{\lambda} = l_{\lambda} \dot{M}_{\ast} \label{eq:LUV}
\end{equation}
where $l_{\lambda}$ is the specific luminosity per unit SFR. We take $l_{\lambda} = l_{1600\angstrom} = 1.08 \times 10^{28} \ \lsfrunits$, which is the result one obtains with the \textsc{bpass} version 1.0 models \citep{Eldridge2009} assuming a constant star formation rate, in which case $l_{1600}$ asymptotes to a constant value on $\sim 100$ Myr timescales, Salpeter stellar initial mass function, a metallicity of $Z = 0.004$, and neglecting binaries and nebular emission. The SFR is assumed to be proportional to MAR, i.e.,
\begin{equation}
    \dot{M}_{\ast}(M_h, z) = \fstar(z, M_h) \dot{M}_b(z, M_h) \label{eq:SFR}
\end{equation}
where the star formation efficiency $\fstar$ encodes all the physics of how galaxy SFRs change as a function of halo mass and potentially redshift.

The key computation in abundance matching is to determine the link between observed luminosity and SFR or halo mass, which is done by requiring the abundance of galaxies with luminosity above $L$ to match the abundance of halos with mass greater than $M_h$, i.e.,
\begin{equation}
	\int_L^{\infty} dL^{\prime} \phi(L^{\prime}) = \int_{M_h}^{\infty} dM_h \frac{dn}{dM_h}
\end{equation}
Here, $\phi(L)$ is some observed galaxy luminosity function (converted from magnitudes to luminosity, perhaps accounting for dust reddening), which we take to be those from \citet{Bouwens2015}. Having determined the mapping between $L$ and $M_h$ at each redshift of interest, one can infer $\fstar$ from Eqs. \ref{eq:LUV} and \ref{eq:SFR}. The remaining challenge is one of extrapolation: given that observed luminosity functions are limited in both redshift and magnitude coverage, one has a few choices:
\begin{itemize}
	\item Commit to the (most likely) Schechter function parameterization inferred in observational studies, and let the evolution in these parameters continue to arbitrarily high redshift. In this case, one need not parameterize $\fstar$ -- its shape will be determined by the interplay between the Schechter function and functional form of $\dot{M}_h(z, M_h)$.
	\item Perform abundance matching only over the range of magnitudes and redshifts covered by observations, and fit a parametric model to $\fstar$ in order to guide extrapolation to fainter luminosities and higher redshifts. For example, adopting a Gaussian or double power-law $\fstar$ then gives one freedom to freely vary the low-mass behavior, and thus faint-end slope of the UVLF.
\end{itemize}
We adopt the latter approach, as in, e.g., \citet{Sun2016}, as it avoids the problem that the faint-end slope of the Schechter function does not properly account for the natural steepening of the halo mass function.

While guaranteed to match observations, and thus a useful tool for making predictions, abundance matching has its drawbacks. For example, we assume the luminosity of galaxies is set by the current SFR alone, with a constant conversion factor between SFR and luminosity. This takes advantage of the fact that the rest-UV emission of galaxies \textit{does} closely track the SFR, as UV emission is dominated by massive, short-lived stars. However, in detail, synthesizing the spectrum properly over all past episodes of star formation can be important, especially for particularly bursty histories. Ignoring the histories of halos also precludes physically-motivated prescriptions for metal and/or dust enrichment, and neglects the diversity of assembly histories of halos at fixed mass and redshift. As a result, we also include two models that treat the evolution of halos in more detail, which we discuss next.

\subsubsection{``Universal'' Semi-Empirical Model (\univ)} \label{sec:univ}
A more sophisticated approach is to essentially run abundance matching in reverse: rather than inferring $\fstar$ from measured UVLFs, and then deciding how to extrapolate to fainter magnitudes and higher redshifts, one chooses a parameterization of $\fstar$ at the outset, and calibrates its free parameters via Markov Chain Monte-Carlo (MCMC) fits to observed luminosities \citep[see, e.g.,][]{Moster2010,Tacchella2013,Behroozi2013}. This approach is essentially a semi-analytic model (SAM), except the efficiency of star formation is parameterized flexibly as a function of halo mass (and perhaps time), rather than setting $\fstar$ by hand via physical arguments, as is done in most SAMs. In this work, we will assume the SFE is universal, i.e., dependent only on $M_h$, hence the name \univ.

We employ two slightly different versions of this model: one in which only the instantaneous MAR is used to model the luminosity of galaxies (\univinst), and another in which the spectrum of all objects is synthesized from their full star formation history (\univhist). In the former case, we use the \citet{Meurer1999} IRX-$\beta$ relation and \citet{Bouwens2014} $\MUV$-$\beta$ relations to correct for dust attenuation, while in the latter case, the $\MUV$-$\beta$ relation is jointly fit with UVLFs, as described in \citet{Mirocha2020a}, and summarized briefly below.

In each model, we assume the star formation efficiency (SFE) is a double power-law (DPL) in $M_h$, i.e.,
\begin{equation}
    \fstar(M_h) = \frac{f_{\ast,10} \ \mathcal{C}_{10}} {\left(\frac{M_h}{M_{\mathrm{p}}} \right)^{-\alphalo} + \left(\frac{M_h}{M_{\mathrm{p}}} \right)^{-\alphahi}} \label{eq:sfe_dpl}
\end{equation}
where $f_{\ast,10}$ is the SFE at $M_h = 10^{10} M_{\odot}$, $M_p$ is the mass at which $\fstar$ peaks, and $\alphahi$ and $\alphalo$ describe the power-law index at masses above and below the peak, respectively. We further assume that the SFR is proportional to the product of $\fstar$ and the baryonic MAR (see \S\ref{sec:halos}). \citet{Mirocha2020a} find best-fit values of $\log_{10} f_{\ast,10} = -1.26$, $\log_{10} (M_{\mathrm{p}}/M_{\odot}) = 11.16$, $\alphalo=0.8$, and $\alphahi=-0.53$.

In the \univinst\ model, the UV luminosity of galaxies is related to their current SFR using a single conversion factor, derived from the \textsc{BPASS} version 1.0 single star models, of $1.08 \times 10^{28} \ \lsfrunits$, which is the $1600\angstrom$ luminosity of a low metallicity, $Z=0.004$, galaxy that has been forming stars at a constant rate for 100 Myr. We convert the resulting luminosities of all halos in the model to magnitudes, and then jointly solve for the amount of UV attenuation, $\AUV$, needed to satisfy the \citet{Meurer1999} IRX-$\beta$ relation and \citet{Bouwens2014} $\MUV$-$\beta$ relations. These dust-corrected magnitudes are then used when fitting the \citet{Bouwens2015} UVLFs.

In the \univhist\ model, the full UV spectrum of all galaxies is synthesized accounting for their entire star formation history. We assume that the metal production rate, $\dot{M}_Z$, is proportional to the SFR, and include dust assuming a dust-to-metal-ratio of $\fdtmr=0.4$\footnote{Note that we hold the stellar metallicity constant, though this has a much smaller effect on colour than dust. See, e.g., Figure 1 in \citet{Mirocha2020a}.}. Then, the dust optical depth is thus given by
\begin{equation}
    \tau_{\nu} = \kappa_{\lambda} N_d = \kappa_{\lambda} \frac{3 \fdtmr M_Z}{4 \pi R_d^2} \label{eq:tau_d}
\end{equation}
where $R_d$ is an effective dust scale length, which we parameterize as a double power-law with free parameters $R_{d,10}$, $M_{\mathrm{p,dust}}$, $\alpha_{\mathrm{lo,dust}}$, and $\alpha_{\mathrm{hi,dust}}$, in analogy with Eq. \ref{eq:sfe_dpl}. The dust opacity is assumed to scale as $\kappa \propto \lambda^{-1}$. The final spectrum of each halo is then computed as the intrinsic spectrum determined from the past star formation history multiplied by $e^{-\tau_{\nu}}$, with $\tau_{\nu}$ determined from Eq. \ref{eq:tau_d}. All magnitudes and UV slopes computed for comparison with observations are generated using the relevant HST filters as a function of redshift \citep[in accordance with the relevant datasets;][]{Bouwens2014,Bouwens2015}.

As in \citet{Mirocha2020a}, we synthesize the spectrum of objects using the entire past star formation history. However, given that simulation snapshots are placed $20$ Myr apart, we further assume a constant SFR rate within each time interval, and perform the spectral synthesis by sub-sampling with timesteps of $\Delta t = 1$ Myr. The best-fit parameters and uncertainties are summarized in full in Table 1 of \citet{Mirocha2020a}.

\subsubsection{Modeling Unresolved Galaxy Populations} \label{sec:URGs}
Because we make no effort in this work to synthesize merger histories for unresolved halo populations (see \S\ref{sec:urhs}), the luminosity of galaxies assumed to be hosted by such halos cannot be modeled in comparable detail to that of galaxies hosted by resolved halos. As a result, in all that follows, the luminosity of unresolved galaxies is tied to halo mass and redshift alone. For example, in \S\ref{sec:reionization} we will show models that employ unresolved source populations but still use the \univhist\ star formation model. In this case, we assume the same exact form for SFE as in the resolved halo case, but use the \citet{Furlanetto2017} analytic model to set halo MAR. The operating assumption of our model is that when averaging over a large number of halos, the individual histories are unimportant as long as the total, population-integrated emissivity is roughly preserved. We explore this assumption in detail in \S\ref{sec:results}.

\subsection{Semi-Numerical Model} \label{sec:seminumerics}
To test the impact of different galaxy modeling choices on reionization, we have developed a custom semi-numerical reionization model. It is similar to, but more rudimentary, than other codes used in the community \citep[e.g.,][]{Mesinger2011,Fialkov2013,Hutter2018}. Because our goal is to gauge the feasibility of the hybrid approach, we do not dwell on our reionization predictions in an absolute sense, but rather focus on the relative differences between direct ($N$-body only) and hybrid ($N$-body augmented with model for unresolved halos; \S\ref{sec:urhs}) techniques.

The fundamental algorithm we use is the excursion set approach, first applied to reionization in \citet{Furlanetto2004}. The underlying argument is simple: a given region is ionized if the number of ionizing photons emitted in that region is equal to or greater than the number of hydrogen atoms,
\begin{equation}
	N_{\gamma}(\mathbf{x},z,R) \geq N_{\mathrm{H}}(\mathbf{x},z,R) , \label{eq:excursion_set}
\end{equation}
where $R$ is the smoothing scale. In order to account for photon originating in neighboring regions, the excursion set algorith is iterative: one begins on large scales and smooths the photon emissivity field with progressively finer top-hat filters until the condition of Eq. \ref{eq:excursion_set} is satified. At this point, one flags either the central cell of the region or the entire region occupied by the smoothing filter. We flag only the central cell as ionized \citep{Mesinger2011,Majumdar2014,Mutch2016}, as this approach seems to most accurately reproduce the results of radiative transfer simulations \citep{Hutter2018}.

Whereas the most efficient codes operate directly on the density field, our models are built on merger trees from $N$-body simulations. As a result, we compute the ionizing photon density directly from the halo field \citep[as in, e.g.,][]{Hutter2018,Hutter2020}, using the galaxy modeling techniques outlined in \ref{sec:galaxies}. In this context, Eq. \ref{eq:excursion_set} takes the form
\begin{equation}
	\sum_i L_{\text{ion}}^i(\mathbf{x},z,R) \geq N_{\mathrm{H}}(\mathbf{x},z,R)
\end{equation}
where $L_{\text{ion}}(M_h)^i$ is the full, forward-modeled ionizing luminosity of the $i^{\text{th}}$ galaxy at location $\mathbf{x}$.

For a fully-sampled halo mass function and 1:1 relationship between halo mass and ionizing luminosity, this expression reduces to the classic condition determined in excursion set theory \citep{Furlanetto2004}, $\fcoll \geq \zeta^{-1}$, where $\zeta$ is the ionizing efficiency $\zeta = \fesc \Nion f_{\ast}$. Our current approach adopts a constant escape fraction $\fesc=0.1$, clumping factor $C=1$, and mean-free path $\Rmfp=10$ Mpc, the latter of which sets the initial smoothing scale used in the excursion set calculation. See, e.g., \citet{Sobacchi2014} for a more sophisticated approach to sub-grid recombinations that eliminates the need for $\Rmfp$.

\section{Results} \label{sec:results}
In this section, we present our main results. First, in \S\ref{sec:results_halos}, we compare our mock galaxy populations to current constraints and explore the extent to which efficient models based on mean halo properties (as in \ares) can be used to populate halo merger trees with galaxies. Then, in \S\ref{sec:reionization}, we turn our attention to reionization, and assess the feasibility of a hybrid approach in which a mix of resolved and unresolved halo populations are used in the model.

\subsection{Galaxy Populations} \label{sec:results_halos}
\begin{figure*}
\begin{center}
\includegraphics[width=0.98\textwidth]{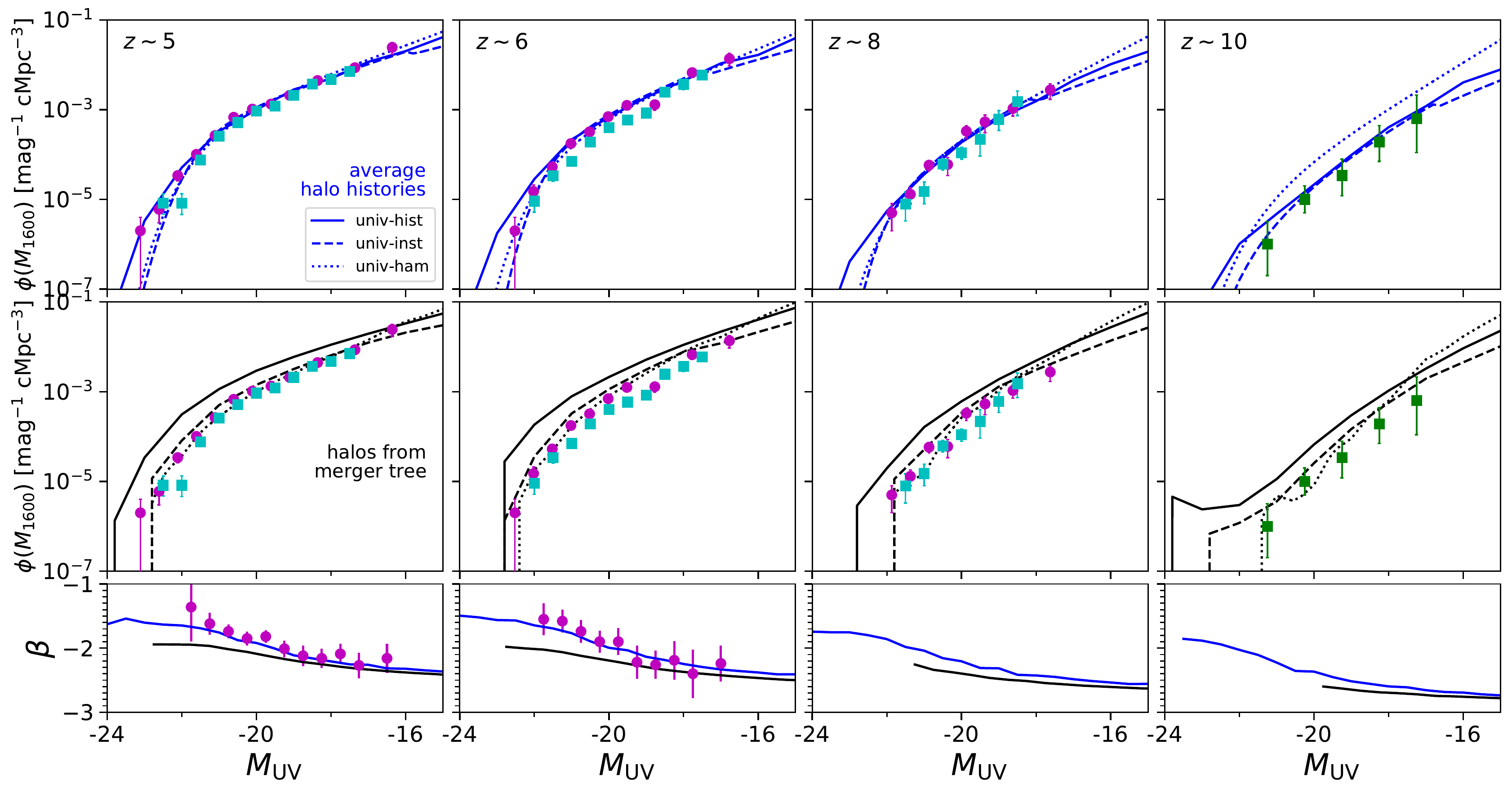}
\caption{{\bf ``Painting'' galaxies into halos from merger trees using the best-fitting parameters derived in idealized models can result in biases} in UVLFs (top, middle) $\MUV$-$\beta$ relations (bottom) from $z \sim 5$ to $z \sim 10$ (left to right). The \univinst\ models (dashed) use a halo's instantaneous SFR to determine its luminosity, while the \univhist\ model (solid) synthesizes the spectrum of each galaxy using its entire star formation history. Both models agree well with the data by construction in the default implementation using idealized halo histories (blue), though applying the calibrated $f_{\ast}$ and $R_d$ curves to a merger-tree-based model (black) results in biases. Measurements shown include UVLFs from \citet{Bouwens2015} (magenta), \citet{Finkelstein2015} (cyan), and \citet{Oesch2018} for $z\sim 10$, and $\MUV$-$\beta$ constraints from \citet{Bouwens2014}. See Table \ref{tab:models} for a summary of the models.}
\label{fig:sam_v_sim}
\end{center}
\end{figure*}

In Figure \ref{fig:sam_v_sim}, we show the UVLFs and UV colours of several galaxy models from $z \sim 5$ to $z \sim 10$. Blue curves in the top row indicate the best-fitting \textsc{ares} models, which are calibrated to the \citet{Bouwens2015} UVLFs at $4 \lesssim z \lesssim 8$ and $\MUV$-$\beta$ relations at $4 \lesssim z \lesssim 6$ from \citet{Bouwens2014}, as described in \citet{Mirocha2020a}. The black curves in the middle row take the best-fitting models for the SFE and dust scale length from the \ares\ models -- which uses idealized halo histories -- and simply ``paints'' them onto halos from the merger tree. In other words, the raw \ares\ models yield preditions for $f_{\ast}$ and $R_d$ as a function of halo mass, and we assume these same scaling relationships hold for halos drawn from a merger tree, even though halos in the merger tree have more complex histories than those in the default \ares\ model.

Three different source models (see Table \ref{tab:models}) are included in Figure \ref{fig:sam_v_sim}: \univhist\ (solid), \univinst\ (dashed), and \ham\ (dotted). These models differ in their treatment of star formation and dust (see \S\ref{sec:galaxies}): \univhist\ synthesizes the luminosity of galaxies using their entire star formation history, while \univinst\ and \ham\ model the luminosity of galaxies as a the product of $f_{\ast}$ and their current MAR only. Regarding dust, \univhist\ adopts the \citet{Mirocha2020a} model, while \univinst\ and \ham\ models correct for dust assuming the \citet{Meurer1999} IRX-$\beta$ relation and $\MUV$-$\beta$ constraints from \citet{Bouwens2014}.

First, we see that the \univinst\ and \ham\ UVLFs barely change when one uses simulated halos rather than idealized halos (compare black and blue dashed curves in top and middle row). This is perhaps unsurprising -- we only use the current MAR to compute galaxy SFRs in these models, and we have shown that the instantaneous MAR of simulated halos agrees well with the simple prescription used in \textsc{ares} (see \S\ref{sec:halos} and Figure \ref{fig:hmar}). However, the results of the \univhist\ model cannot be so simply transplanted. When ``painted'' onto halos from a merger tree, a systematic bias emerges in the direction of brighter galaxies and bluer colours (solid black lines, middle and bottom row). This suggests that the increased complexity of halos drawn from a merger tree can have a discernible impact on observables. This effect is relatively small at $8 \lesssim z \lesssim 10$, though at $z \lesssim 8$, the UVLF of merger-tree-generated galaxies is systematically brighter by $\sim 1$ magnitude, while UV colours are bluer by $\Delta \beta \sim 0.2-0.5$, than the corresponding model using idealized halos in \ares.

The emergence of biases here may not come as a surprise; one cannot necessarily expect a model calibrated with an idealized set of halos to work when applied to a population of halos generated in $N$-body simulations. Had it been the case that \textsc{ares} models can be transplanted onto simulated halos as-is for more detailed 3-D predictions, that would have been exceedingly convenient. One could, e.g., perform parameter inference in the much more efficient \textsc{ares} framework and only turn to simulations when galaxy survey or 21-cm mocks were needed. Instead, we must modify the \textsc{ares} models before proceeding if the $N$-body-based models are to remain in good agreement with UVLFs and colours. We will revisit such modifications momentarily.


\begin{figure*}
\begin{center}
\includegraphics[width=0.98\textwidth]{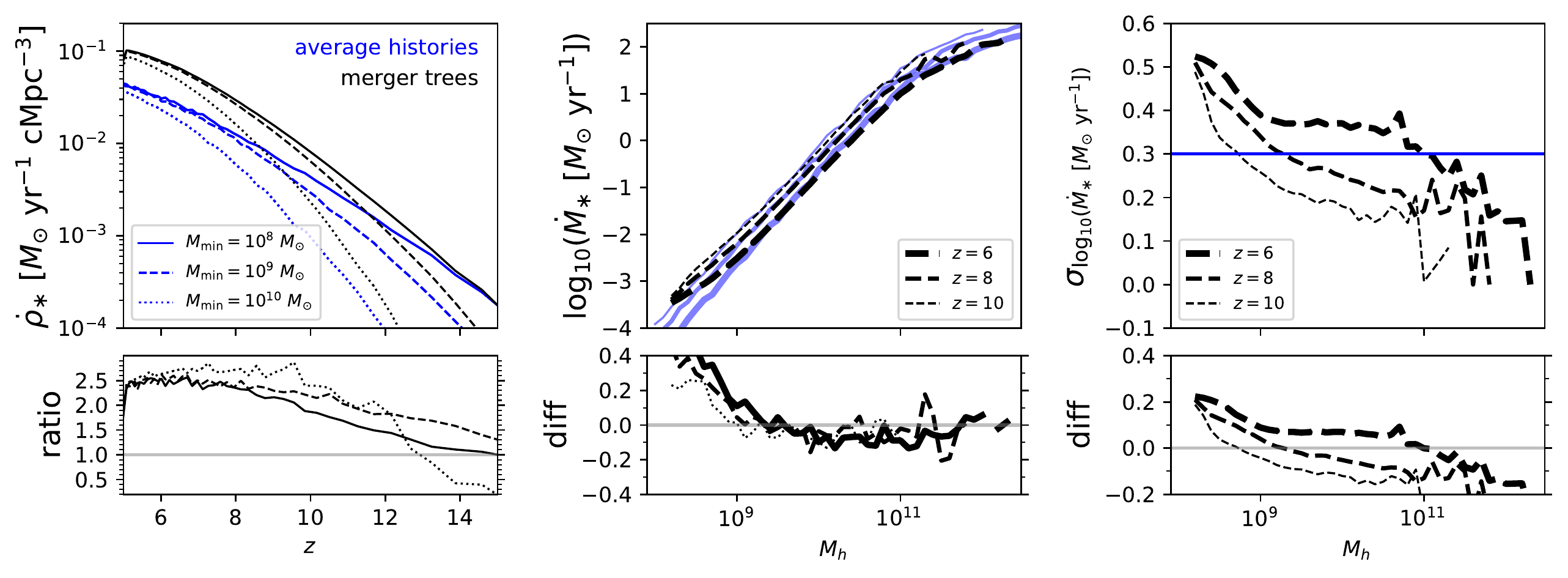}
\caption{{\bf The SFRD (left), $M_h$-SFR relation (center), and scatter in the $M_h$-SFR relation for models with (black) and without (blue) merger trees.} Here, we focus on the \univhist\ source model, and show differences between the idealized and merger tree models in the bottom row. The difference in the amount of star formation in the merger tree models is driven both by additional scatter not present in the idealized models, as well as the poorly-resolved MAR histories of low-mass halos, particularly at late times when mass accretion rates are lower. During reionization, $6 \lesssim z \lesssim 10$, the offset in massive halos $M_{\min} = 10^{10} \ M_{\odot}$ is $\sim 2$x (lower left) results in a nearly constant $\sim 2$x offset in massive halos (dotted line, lower left panel), while halos with $M_h \lesssim 10^{10} \ M_{\odot}$ show SFRs up to $\sim 0.5$ dex higher in the merger tree model at $z \sim 6$ (lower right).}
\label{fig:sfrd_sfr}
\end{center}
\end{figure*}

In order to identify the origin of the systematic biases illustrated in Figure \ref{fig:sam_v_sim}, in Figure \ref{fig:sfrd_sfr}, we explore the star formation rate density (SFRD) and SFR as a function of halo mass for each model, varying $\Mmin$ from $10^8$ to $10^{10} \ M_{\odot}$ to isolate any potential mass-dependent effects. Focusing first on the left-most panel, we see that the SFRD is systematically offset between merger-tree-based models and the default \ares\ models. As shown in the bottom panel, this offset is a factor of $\sim 2.5-3$. The offset is comparable regardless of the minimum mass threshold, suggesting at most a mild mass-dependence. This suggests that the systematic offset in the UVLFs (Figure \ref{fig:sam_v_sim}) is at least in part due to elevated SFR levels in halos from merger trees. A systematic under-estimate of dust reddening is likely also at work, given the bluer colors in merger-tree-based models (bottom row of Figure \ref{fig:sam_v_sim}). We revisit this point momentarily.

The middle panel of Figure \ref{fig:sfrd_sfr} shows that in all halos above $M_h \sim 10^{10} \ M_{\odot}$, the relationship between halo mass and \textit{mean} SFR is in good agreement between merger tree and idealized models, apart from a $\sim 0.2$ dex bias at $M_h \sim 10^9 \ M_{\odot}$ at $z=6$. This partially explains the bias in the cosmic SFRD, and is likely due to the elevated MAR inferred from simulated halos relative to the \citet{Furlanetto2017} model. This does not appear to be entirely a resolution artifact (see Appendix \ref{sec:convergence}). The scatter in the SFR is also higher in the merger tree galaxies (right panel of Figure \ref{fig:sfrd_sfr}), which also contributes to the SFRD bias given that the scatter is log-normal. This could be, at least in part, a resolution effect (see Appendix \ref{sec:convergence}), particularly at $M_h \lesssim 10^9 \Msun$.

Clearly, the best-fitting parameters of models built on idealized halo populations require some revision before use in a model based on merger trees. A reduction in the normalization of the star formation efficiency, $f_{\ast,10}$, is certainly warranted, as is a reduction in the dust scale length, $R_{d,10}$ or boost in dust yield, $f_d$. Adjusting the dust parameters will compensate for the blueward bias in UV colours for merger tree models, which has two causes: (i) the steeper-on-average growth histories of simulated halos relative to idealized growth histories mean that in a given halo mass bin, galaxies have (on average) less stellar mass, and thus dust mass, and (ii) the additional scatter in simulated halo populations means that there is more scatter in dust mass at fixed halo mass (or MAR) -- scatter has an asymmetric effect on the $\MUV$-$\beta$ relation, since over-luminous or under-dusty galaxies willways outnumber `typical' galaxies in the $\MUV$ bin they up-scatter into \citep[see \S3.2 in][]{Mirocha2020a}.

Given infinite computational resources, we would simply run the model calibration using simulated halos from the outset. However, this is computationally very demanding, e.g., our 80 Mpc box has $\sim 10^7$ halos at $z \sim 6$ (not counting all progenitors), whereas the default approach in \ares\ evolves $\sim 10^3$ ``halos'' representative of their mass. There is thus a need to find a way to map idealized models onto more sophisticated halo populations, especially given the impact such choices have on predictions for reionization (as we discuss next in \S\ref{sec:reionization}).

\begin{figure*}
\begin{center}
\includegraphics[width=0.98\textwidth]{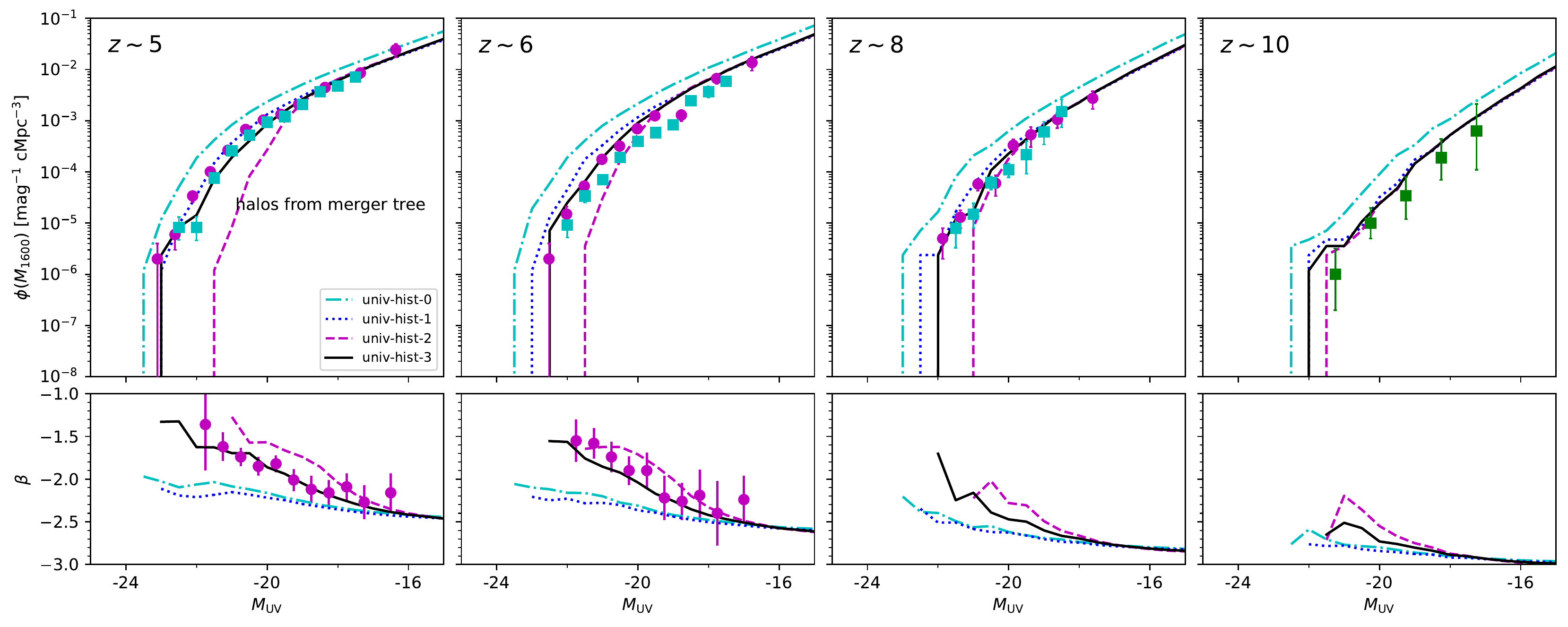}
\caption{{\bf Simple modifications to the idealized \univhist\ models can recover agreement with UVLFs and colours.} Same as Figure \ref{fig:sam_v_sim}, except each model uses halos from merger tree. Model variants 1-4 (indicated in legend) multiply the normalization of $f_{\ast,10}$ by 0.4 and $R_{d,10}$ by $\{1,0.8,0.8,0.7\}$, respectively. Variants 3 and 4 also take $\alphahi=-0.2$ and $-0.1$, respectively, and $(\alphaloD=0.45, \alphahi=0.55)$. See text for more discussion of these adjustments.}
\label{fig:kludges}
\end{center}
\end{figure*}

In Figure \ref{fig:kludges}, we explore several modifications to the default \ares\ models, designed to reduce tension with observations when employing merger trees from $N$-body simulations. First, we simply re-scale the normalization of the SFE, $f_{\ast,10}$ by a factor of $0.4$ (dotted cyan), which reduces tension with UVLFs at the faint-end but exacerbates tension in $\MUV$-$\beta$. So, in variants 2-4, we adjust also the dust scale length $R_{d,10}$ by factors of $\{0.8,0.8,0.7\}$, which helps with $\MUV$-$\beta$ and the bright-end of the UVLF. After some experimentation, we find that changes to the slope of the relationship between dust scale length and halo mass is also required. Whereas the best-fit values determined in \citet{Mirocha2020a} are $\alphaloD=0.7$ and $\alphahiD=0.1$; here, we find that a relation closer to a single power-law with $\alphalo \simeq \alphahiD \simeq 0.5$ minimizes tension in UVLFs and UV colours (variants 3 and 4 in Figure \ref{fig:kludges}). We also find that a shallower relation between $f_{\ast}$ and halo mass above the SFE peak is required (solid black curves). For the rest of the paper, we thus adopt a model with $\alphaloD=0.45$, $\alphahiD=0.55$, and $\alphahi=-0.2$. We defer a more thorough exploration of such ``kludges'' to future work, as it will likely continue to be an important step in mapping best-fitting idealized models to more sophisticated 3-d simulations\footnote{Note that there is a degeneracy between the efficiency of star formation, dust scale length, and duty cycle of star formation \citep{Mirocha2020b} that we ignore here for simplicity.}. In the remainder of this work, we adopt variant \#3 of the \univhist\ model shown in Figure \ref{fig:kludges}.

\subsection{Reionization} \label{sec:reionization}
With a calibrated source model in hand, we now turn our attention to predictions for reionization. In this work, we are uninterested in reionization predictions in an absolute sense -- instead, we seek to assess the feasibility of models that only treat in detail the halos hosting directly-observable high-$z$ galaxies. If sufficiently accurate, such hybrid models can be deemed fit for future studies, particularly those for which a more detailed picture of bright galaxies during reionization is warranted.

\begin{figure*}
\begin{center}
\includegraphics[width=0.98\textwidth]{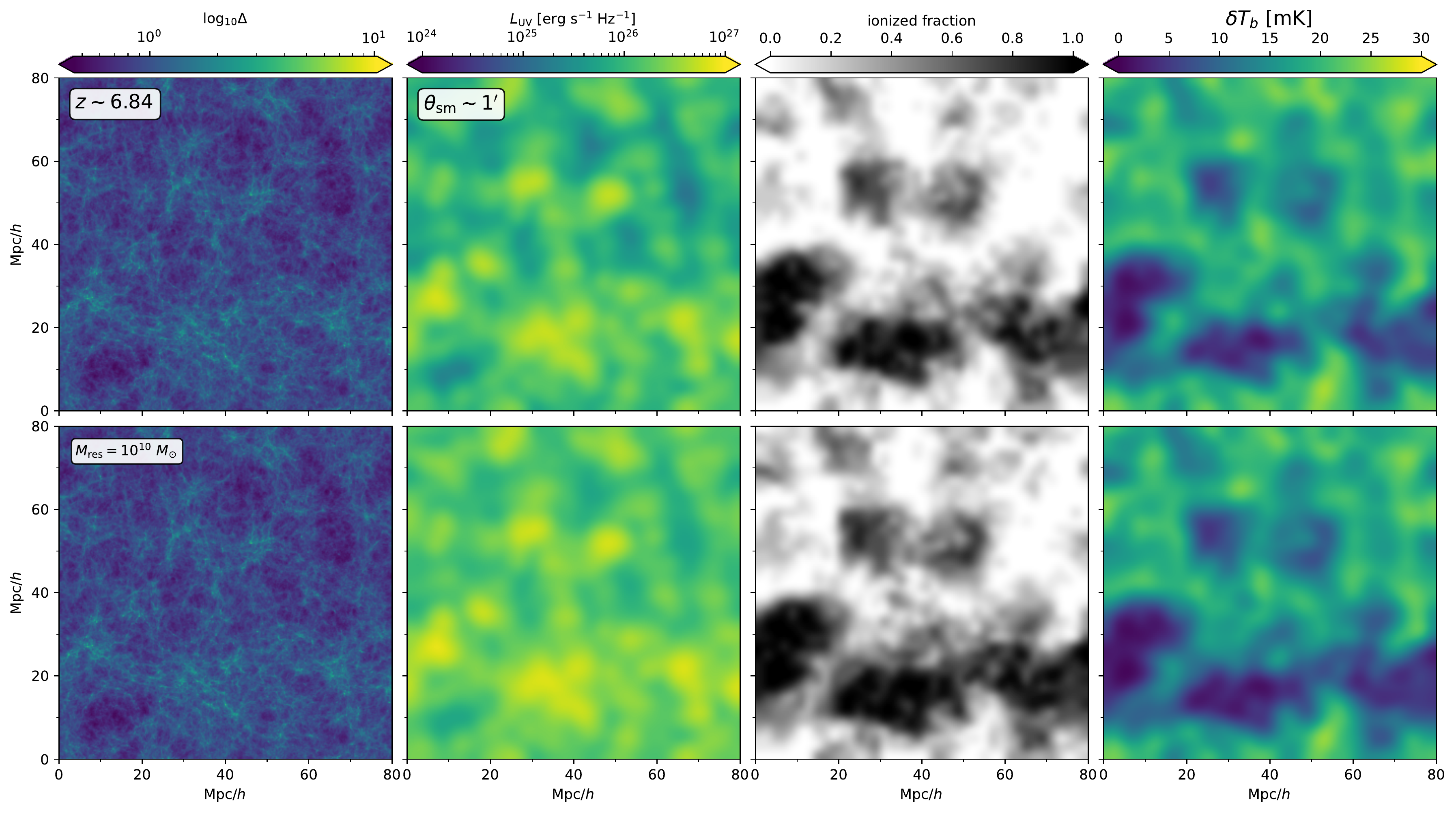}
\caption{{\bf Summary of model predictions for direct (top) and hybrid (bottom) methods.} From left to right, we show the cosmic density field, rest 1600\angstrom\ luminosity field smoothed to arcminute resolution, the ionization field, and the 21-cm brightness temperature field. All panels are $20 \ \Mpch$ thick projections along the line of sight. The volume-averaged ionized fraction at $z \simeq 6.84$ is $Q_{\mathrm{ion}} \simeq 0.5$ in this model. The model in the bottom row is hybrid, in the sense that halos with $M_h < 10^{10} \Msun$ are modeled in aggregate following \S\ref{sec:urhs}.}
\label{fig:fields_summary}
\end{center}
\end{figure*}

We begin in Figure \ref{fig:fields_summary} with an illustration of our model's predictions for several fields of interest, both for direct ($N$-body halos only; top row) and hybrid ($N$-body augmented with model for unresolved halos, see \S\ref{sec:urhs}; bottom row) models. From left to right, we show narrow $20$ Mpc projections through the cosmic density field, rest-UV luminosity field, ionization field, and 21-cm brightness temperature field. Each field (apart from the raw density field in the left panel) is computed on a coarse grid with $(2 \ h^{-1} \ \mathrm{Mpc})^3$ voxels, and is smoothed using a bicubic interpolation only for aesthetic purposes, resulting in maps with $\sim 1$ arcminute resolution.

The bottom row of Figure \ref{fig:fields_summary} adopts a hybrid approach, in which halos below a resolution limit of $\Mres = 10^{10} \ M_{\odot}$ are modeled in aggregate, rather than directly from the merger tree. Qualitatively, the agreement is very good, though there are quantitative differences. For example, dark spots in the hybrid luminosity map are darker than the direct model, while the ionization field has slightly less small-scale structure as well.

\begin{figure}
\begin{center}
\includegraphics[width=0.49\textwidth]{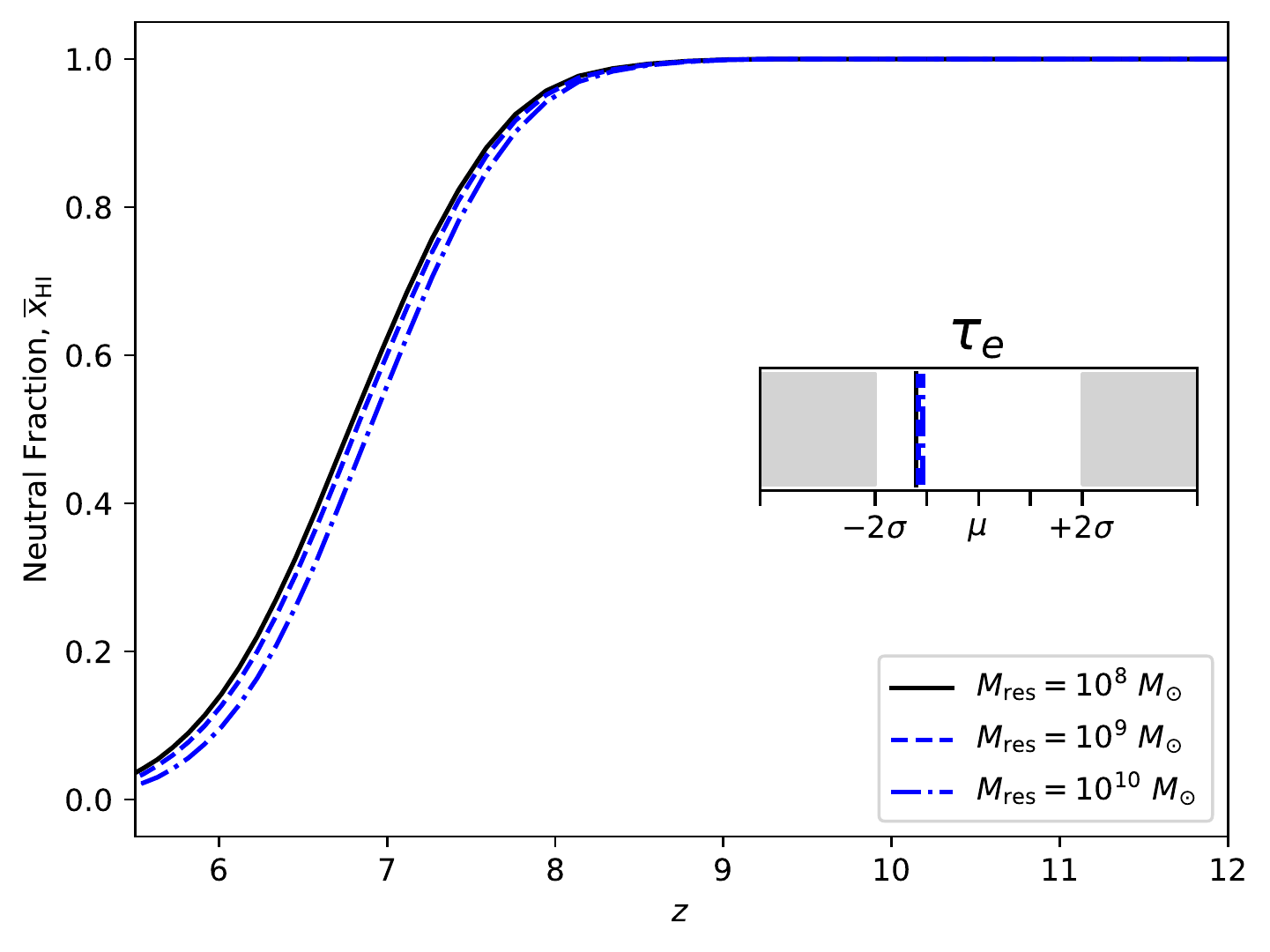}
\caption{{\bf Predictions for the mean reionization history and CMB optical depth, $\tau_e$ (inset), for direct (black) and hybrid (blue) models.} Dashed and dot-dashed blue curves indicate increasingly aggressive hybrid modeling, with halos less massive than $\Mres = 10^9$ and $10^{10} \ \Msun$, respectively, treated in aggregate.}
\label{fig:rei_mean}
\end{center}
\end{figure}

\begin{figure*}
\begin{center}
\includegraphics[width=0.98\textwidth]{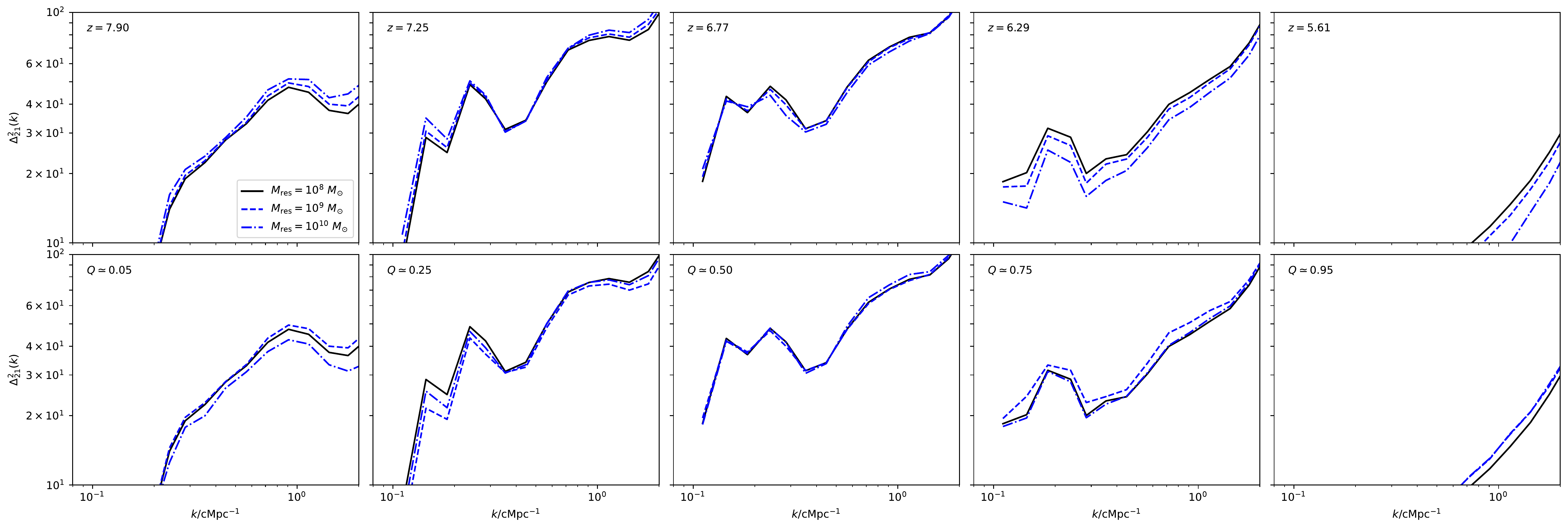}
\caption{{\bf Predictions for the dimensionless 3-D 21-cm power spectrum.} From left to right, we compare direct (black) and hybrid (blue) methods at five different stages of reionization. The top row shows power spectra at a given redshift, while the bottom row compares power spectra at fixed volume-averaged ionized fraction, as indicated by the annotations in the upper-left corner of each panel.}
\label{fig:rei_ps21}
\end{center}
\end{figure*}

In Figures \ref{fig:rei_mean} and \ref{fig:rei_ps21}, we examine more quantitatively the differences between the reionization predictions of direct and hybrid models. First, in Figure \ref{fig:rei_mean}, we focus on the mean reionization history and Thomson scattering optical depth from \textit{Planck} (inset shows $2\sigma$ credibility range for $\tau_e = 0.055 \pm 0.009$ constraint). The solid black line is our fiducial `direct' model, in which no model for unresolved halos is adopted. Dashed and dotted blue lines employ unresolved halos below $10^9$ and $10^{10} \ M_{\odot}$, respectively. The magnitude of the difference between models is present, as expected, but very small. Reionization starts slightly earlier in the hybrid models, though this amounts to a negligible difference in $\tau_e$ (inset). The sense of this offset is expected given that unresolved halos, by construction, follow a \citetalias{PressSchechter1974} mass function, and are thus more abundant than the simulated halos, which follow \citet{Tinker2010} more closely, and suffer from some incompleteness at low mass. We have made no effort to adjust the galaxy parameters differentially between resolved and unresolved halo populations due to the good initial agreement shown here, but note that using the \univinst\ or \ham\ models for unresolved halos could provide an even better model.

In Figure \ref{fig:rei_ps21}, we show the 3-d power spectrum of the 21-cm background, $\Delta_{21}^2(k) \equiv k^3 P_{21}(k) / 2\pi^2$, assuming the IGM is fully saturated, i.e., the spin temperature greatly exceeds the CMB temperature in neutral regions. The 21-cm power spectrum could differ in hybrid models even if the mean history is preserved exactly, so we compare the results of direct (black) and hybrid (blue) schemes at five different stages of reionization, both at fixed redshift (top row) and ionized fractions (bottom row). Differences are most noticeable late in reionization (right-most panels), but generally small $\sim 20-30$\% when the volume-averaged ionized fraction is under $\sim 0.75$. This suggests that, on the scales accessible to current experiments ($k \lesssim 1 \ \Mpch$), treating faint galaxies in aggregate has little effect on the power spectrum. This is consistent with the good apparent agreement in ionization field morphology shown in Figure \ref{fig:fields_summary}, though the small box size effects are readily apparent in the power spectra on large $k \lesssim 0.3 \ \Mpchinv$ scales.

Finally, we turn our attention from the potential shortcomings of the hybrid approach to its advantages. We have already seen how neglecting the detailed histories of galaxies can bias models calibrated to high-$z$ UVLFs (see Figure \ref{fig:sam_v_sim}), but have yet to explore these mock galaxy populations in any further detail. Now, we flag galaxies bright enough to be detected in upcoming surveys, including a $\mAB \simeq 29.3$ cut \citep[as in][]{Mason2015}, comparable to the JADES deep \citep{Rieke2019} and CEERS \citep{Finkelstein2017} blank field surveys, as well as a shallower $\mAB \simeq 26.5$ cut to mimic the high-latitude survey (HLS) of the \textit{Nancy Grace Roman Space Telescope} \citep{Spergel2015,Dore2019}. Note that our $80 \ \Mpchinv$ box corresponds to a $\sim 2000$ arcmin$^2$ field of view at $z \sim 7$, whereas planned deep surveys target much smaller areas of $\lesssim 100$ arcmin$^2$.

\begin{figure}
\begin{center}
\includegraphics[width=0.49\textwidth]{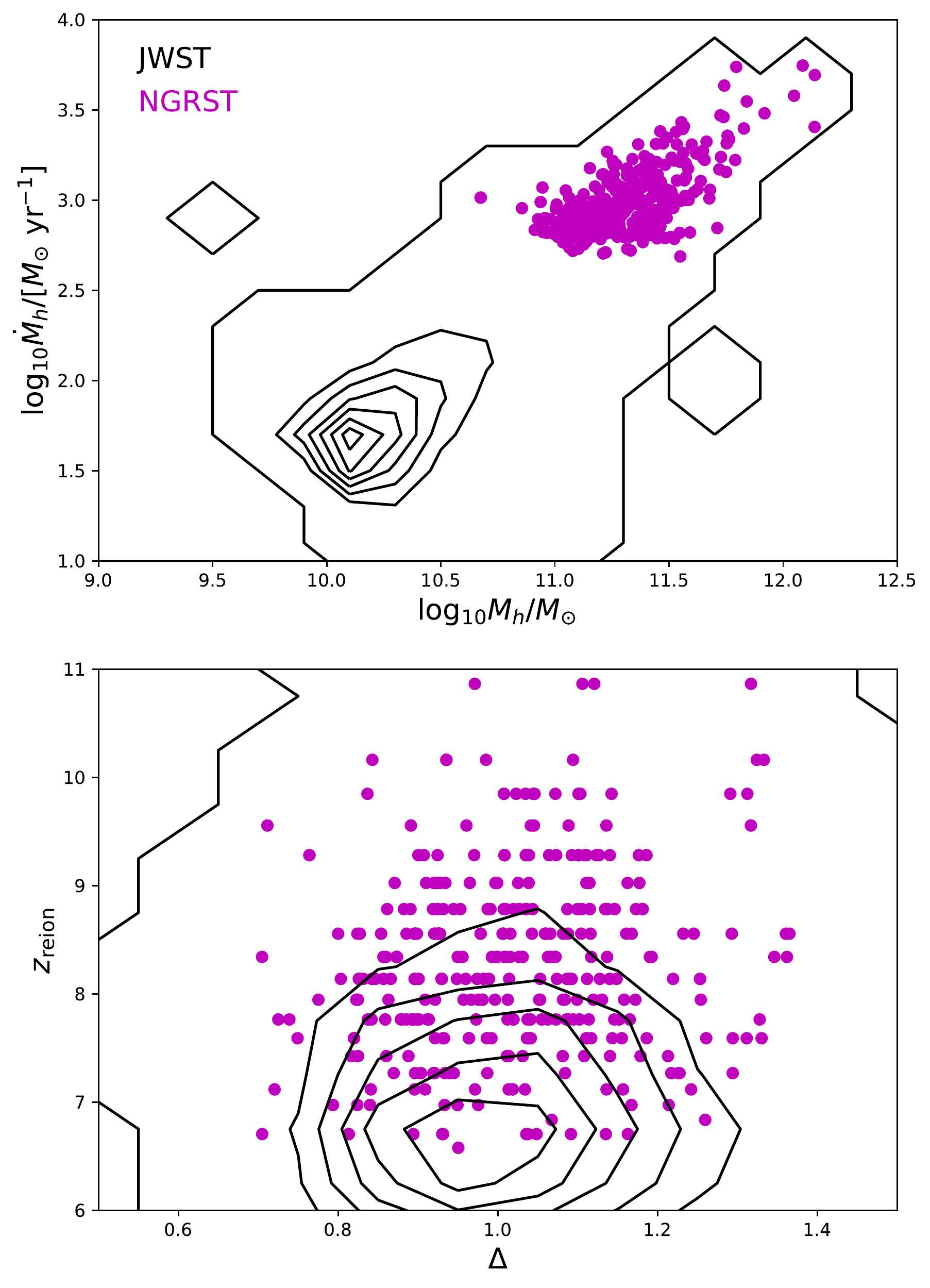}
\caption{{\bf Halo and environmental properties of galaxies detectable by JWST wide-field survey and \textit{Roman} HLS.} \textit{Top:} Distribution in halo mass -- MAR space for JWST (black contours) and HLS (magenta) sources. \textit{Bottom:} Distribution in large-scale environment, quantified by density $\Delta$ (relative to cosmic mean) smoothed with $8 \ \Mpchinv$ spherical top-hat, and reionization redshift, $\zreion$, of parent voxel.}
\label{fig:survey_hists}
\end{center}
\end{figure}

In Figure \ref{fig:survey_hists}, we first show joint distribution of halo mass vs. MAR of halos hosting detectable galaxies in each of these surveys at $z \sim 7$, as well as the large-scale density and reionization redshift of these galaxies, both averaged in $8$ Mpc spherical top-hat windows. In the top panel, we see that a JWST wide-field can detect galaxies in halos as small as $\sim 10^{10} \ \Msun$. The most massive detectable halos are $\sim \times 10^{12} \ \Msun$, while 68\% of JWST sources lie in the $5 \times 10^{10} \lesssim M_h / \Msun \lesssim 10^{11}$ range. \textit{Roman} sources (magenta) are more massive on average, as expected, by roughly an order of magnitude, and have mass accretion rates $\gtrsim 10^3 \MARunits$. In the bottom panel of Figure \ref{fig:survey_hists}, we show the large-scale density and ionization environments of each galaxy population. While all \textit{Roman} galaxies live in massive, highly-accreting halos, they live in fairly typical density and ionization environments.

\begin{figure*}
\begin{center}
\includegraphics[width=0.98\textwidth]{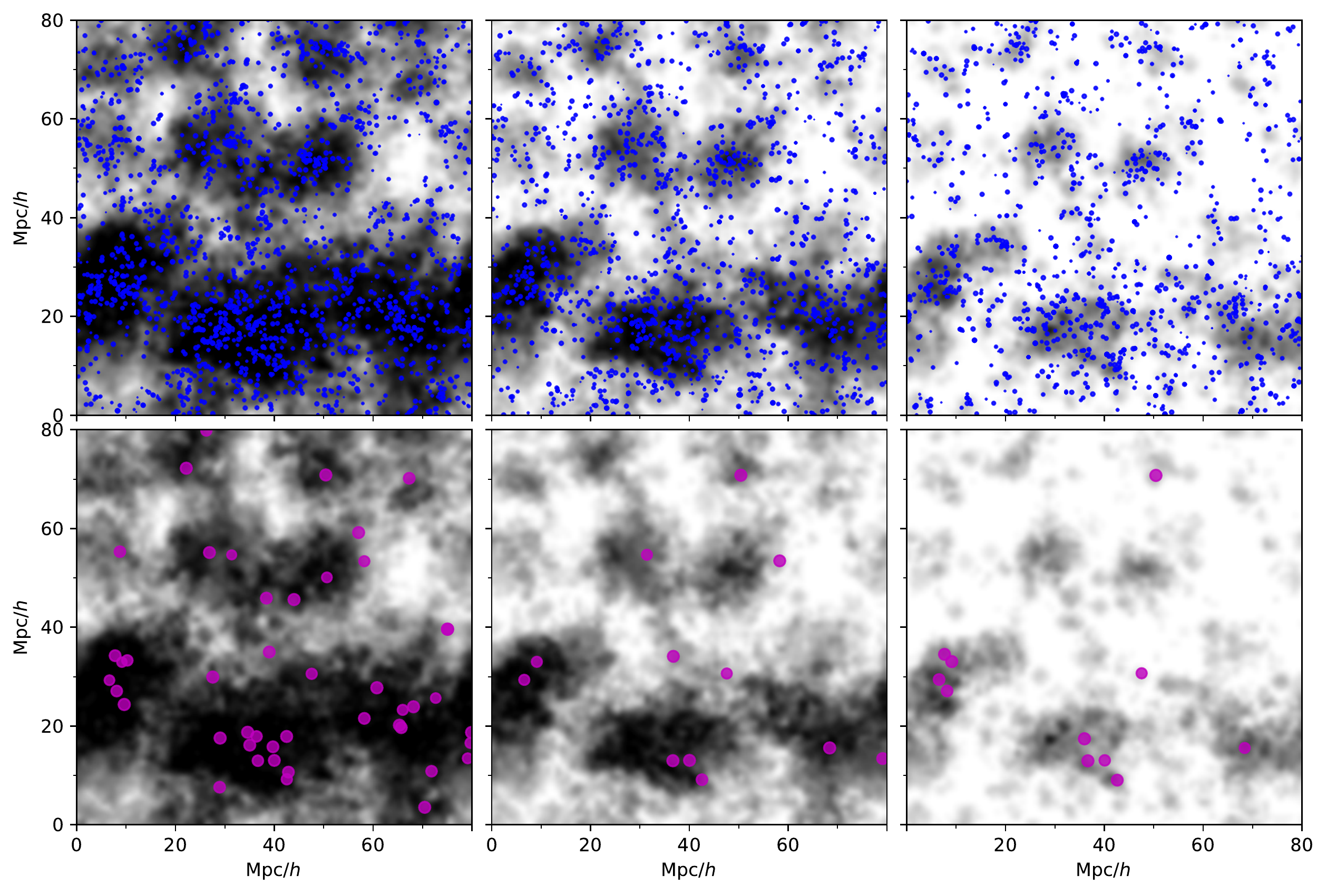}
\caption{{\bf Alignment of 'detectable' galaxies at three different ionized fractions, $Q=0.75$, $Q=0.5$, and $Q=0.25$, from left to right.} Blue points in the top row indicate galaxies bright enough to be detected in a JWST wide-field survey ($\mAB \lesssim 29.3$), while magenta points in the bottom row indicate galaxies bright enough to be detected in the \textit{Roman} HLS ($\mAB \lesssim 26.5$). Plot symbol sizes in each panel increase linearly relative to the limiting magnitude of the survey, $|m_{\rm AB,lim} - \mAB|$. Note that  $80 \ \Mpchinv$ subtends an angle of $\sim 46$ arcmin at $z \sim 7$, and that each of these images are narrow 20 cMpc projections along the line of sight.}
\label{fig:survey}
\end{center}
\end{figure*}

Next, in Figure \ref{fig:survey}, we show these mock galaxy populations overlaid on top of the ionization field. Focusing first on the top panel, we immediately see the advantage of depth -- there are $\sim 50$x more galaxies in the field that are detectable with JWST than there are for the HLS\footnote{Note that we have not performed selection based on color. `Detectable` here simply means that a galaxy is bright enough to be detected in the filter closest to rest $1600 \angstrom$.}. Similarly, it is unsurprising to find groups of these galaxies in ionized regions, with fewer visible in the still-neutral regions of the IGM. However, the same is not always true of the \textit{Roman} sources. As the brightest galaxies in the field, one might naively expect them to always lie at the center of large ionized bubbles, yet they often lie on the outskirts, or in much smaller, partially neutral regions.

This offset between \textit{Roman} sources and ionized bubbles suggests that halos experiencing the strongest fluctuations in MAR are often not in the densest regions of the simulation volume. The fact that \textit{Roman} sources live in all density and ionization environments, as shown in Figure \ref{fig:survey_hists}, which examined the whole volume rather than a narrow projection, shows that this is not a projection effect. This result has important implications for efforts to cross-correlate galaxy surveys and 21-cm maps, which will seek a statistical detection of the expected anti-correlation between galaxies and 21-cm emission. Such cross correlations are likely only viable for the wide -- but shallow -- fields accessible with \textit{Roman}, and thus susceptible to any effect which changes the locations of the brightest galaxies with respect to the most ionized regions of the Universe. Adding context to deep JWST surveys covering much narrower fields than each panel in Figure \ref{fig:survey} with 21-cm observations is challenging, but possible \citep{Beardsley2015}.

\section{Discussion \& Conclusions} \label{sec:conclusions}
We have developed a semi-numerical model of reionization anchored to the outputs of $N$-body simulations, in the spirit of other recent works focused on implementing more detailed galaxy models in efficient reionization models \citep[e.g.,][]{Mutch2016,Hutter2020}. Though this approach is slower than the classic semi-numeric approach \citep{Mesinger2011,Fialkov2013}, in which ionizing photon production is linked directly to the linearly-evolved cosmic density field, it allows a more detailed treatment of galaxy formation using semi-analytic modeling techniques, and thus may be warranted for applications in galaxy -- 21-cm cross correlations or contextualizing galaxy surveys with 21-cm observations more broadly.

Our main goal was to determine the extent to which galaxy models tied to $N$-body merger trees differ from simpler approaches that do not rely on simulations, as the latter cannot treat individual galaxy formation histories in detail. Additionally, we explored the prospects of a `hybrid` approach, in which low-mass halos are modeled only in aggregrate. This is likely more efficient than other hybrid techniques, many of which aim to extend merger trees beyond their native resolution limits using Monte Carlo techniques \citep[e.g.,][]{Nasirudin2020,Qiu2021} and thus aim to keep a similar level of detail in resolved and unresolved halo populations, in contrast to our approach. Though our approach of course requires relatively crude $\gtrsim 2$ Mpc grid resolution in order to justify a well-sampled halo mass function at the low-mass end, current observations are unlikely to breach these scales. As a result, a hybrid approach such as ours is most well-suited to current and near-future observations, limited to large-scale 21-cm fluctuations and the brightest galaxies accessible to surveys. It also may provide a means to include reionization self-consistently in model lightcones that extend to $z = 0$ for line intensity mapping applications \citep[e.g.,][]{Yang2020}, which generally cannot afford to resolve halos with $M_h \lesssim 10^{10} \ \Msun$ and run to $z=0$.

The key findings of our work are as follows:
\begin{itemize}
  \item Simple models, in which idealized halo histories are used instead of merger trees, are biased with respect to models based on merger trees at the level of $\sim 1$ magnitude in UVLFs at $6 \lesssim z \lesssim 8$ (see Figure \ref{fig:sam_v_sim}). This is due to the increased diversity of simulated halo growth histories, which affects the range of SFRs and dust reddening present in any given halo mass (or MAR) bin (see Figure \ref{fig:hhist} and \ref{fig:sfrd_sfr}).
  \item However, biases inherent to the simpler models can largely be eliminated by tuning nuisance parameters like the overall normalization of the SFE or dust scale length (see Figure \ref{fig:kludges}). In other words, the more efficient  inference approach based on mean halo histories (as in \ares\ and \citet{Furlanetto2017}) is still a viable technique to calibrate galaxy models, meaning more expensive MCMCs that use merger trees constructed from $N$-body simulations can generally be avoided.
  \item Given the advantages of modeling bright galaxies in detail, we explore the extent to which the faintest galaxies can be modeled in aggregate, allowing most computational resources to be diverted to detailed modeling of the galaxies bright enough to be detected in upcoming surveys. We find that galaxies in halos just beyond the reach of deep surveys, $M_h \lesssim 10^{10} \ \Msun$, can be modeled in aggregate at little cost -- a small bias in the mean reionization history and $\sim 20-30$\% offsets in the 21-cm power spectrum, likely comparable to or smaller than systematic uncertainties in the modeling \citep[see, e.g.,][]{Schneider2020,Mirocha2021a}.
  \item The offset between hybrid and direct semi-numeric models are, at least in part, due to differences in the mean MAR of halos $M_h \lesssim 10^{10} \ \Msun$ in our $N$-body simulations compared to analytic models for halo MAR (see Figure \ref{fig:hmar}). This does not appear to be entirely a resolution artifact (see Appendix \ref{sec:convergence}). As a result, modification of the pure power-law decline in MAR at low $M_h$ in the future may bring direct and hybrid results for the reionization history into closer agreement. Corrections to the unresolved HMF and/or galaxy parameters may also be warranted.
  \item Bright galaxies detectable with \textit{Roman} are those with the highest accretion rates (see Figure \ref{fig:survey_hists}), but do not always reside in the densest most ionized regions of the volume (Figure \ref{fig:survey}). This may complicate cross-correlation efforts if limited to bright galaxy samples. However, it may also serve as a test of models of inflow-driven star formation. We defer a more detailed investigation of this possibility to future work.
\end{itemize}

\section*{Acknowledgments}
The authors thank Brad Greig for helpful feedback on this work, as well as the anonymous referee for many useful suggestions that helped improve the paper. J.M. acknowledges the Canadian Institute for Theoretical Astrophysics for support through a CITA National Fellowship. A.L. acknowledges support from the New Frontiers in Research Fund Exploration grant program, a Natural Sciences and Engineering Research Council of Canada (NSERC) Discovery Grant and a Discovery Launch Supplement, a Fonds de recherche Nature et technologies Quebec New Academics grant, the Sloan Research Fellowship, the William Dawson Scholarship at McGill, as well as the Canadian Institute for Advanced Research (CIFAR) Azrieli Global Scholars program. This material is based upon work supported by the National Science Foundation under Grant No. 1636646, the Gordon and Betty
  Moore Foundation, and institutional support from the HERA collaboration
  partners. HERA is hosted by the South African Radio Astronomy Observatory,
  which is a facility of the National Research Foundation, an agency of the
  Department of Science and Technology. This work used the Extreme
  Science and Engineering Discovery Environment (XSEDE), which is supported by
  National Science Foundation grant number ACI-1548562
  \citep{xsede2014}. Specifically, it used the Bridges system, which is
  supported by NSF award number ACI-1445606, at the Pittsburgh Supercomputing
  Center \citep{bridges2015}. Additional computations were made on the supercomputer Cedar at Simon Fraser University managed by Compute Canada. The operation of this supercomputer is funded by the Canada Foundation for Innovation (CFI).The authors would also like to thank the Aspen Center for Physics for hosting \textit{Cosmological Signals from Cosmic Dawn to the Present} in 2018, where this work was initiated.

\textit{Software:} numpy \citep{numpy}, scipy \citep{scipy}, matplotlib \citep{matplotlib}, h5py\footnote{\url{http://www.h5py.org/}}, mpi4py \citep{mpi4py1}, and \textsc{powerbox}\footnote{\url{https://github.com/steven-murray/powerbox}} \citep{Murray2018}.

\textit{Data Availability:} The simulation data presented in this article is available upon request.

\bibliography{references}
\bibliographystyle{mn2e_short}

\appendix

\section{Convergence Study} \label{sec:convergence}
As described in \S\ref{sec:rhs}, at redshifts $z \lesssim 8$, the mass accretion rates (MARs) of low-mass halos are poorly-resolved, even though they may be well-resolved in mass. This is due to the discrete nature of the matter field and $20$ Myr cadence between snapshots, effectively resulting in shot noise when the expected number of particles accreted per timestep is small. One approach, which we employ in the main text, is to simply smooth the MAR of halos over 5 snapshots (100 Myr), to average out this effect. However, to test the validity of this approach, we must compare the smoothed MAR to the results of higher resolution simulations.

\begin{figure*}
\begin{center}
\includegraphics[width=0.98\textwidth]{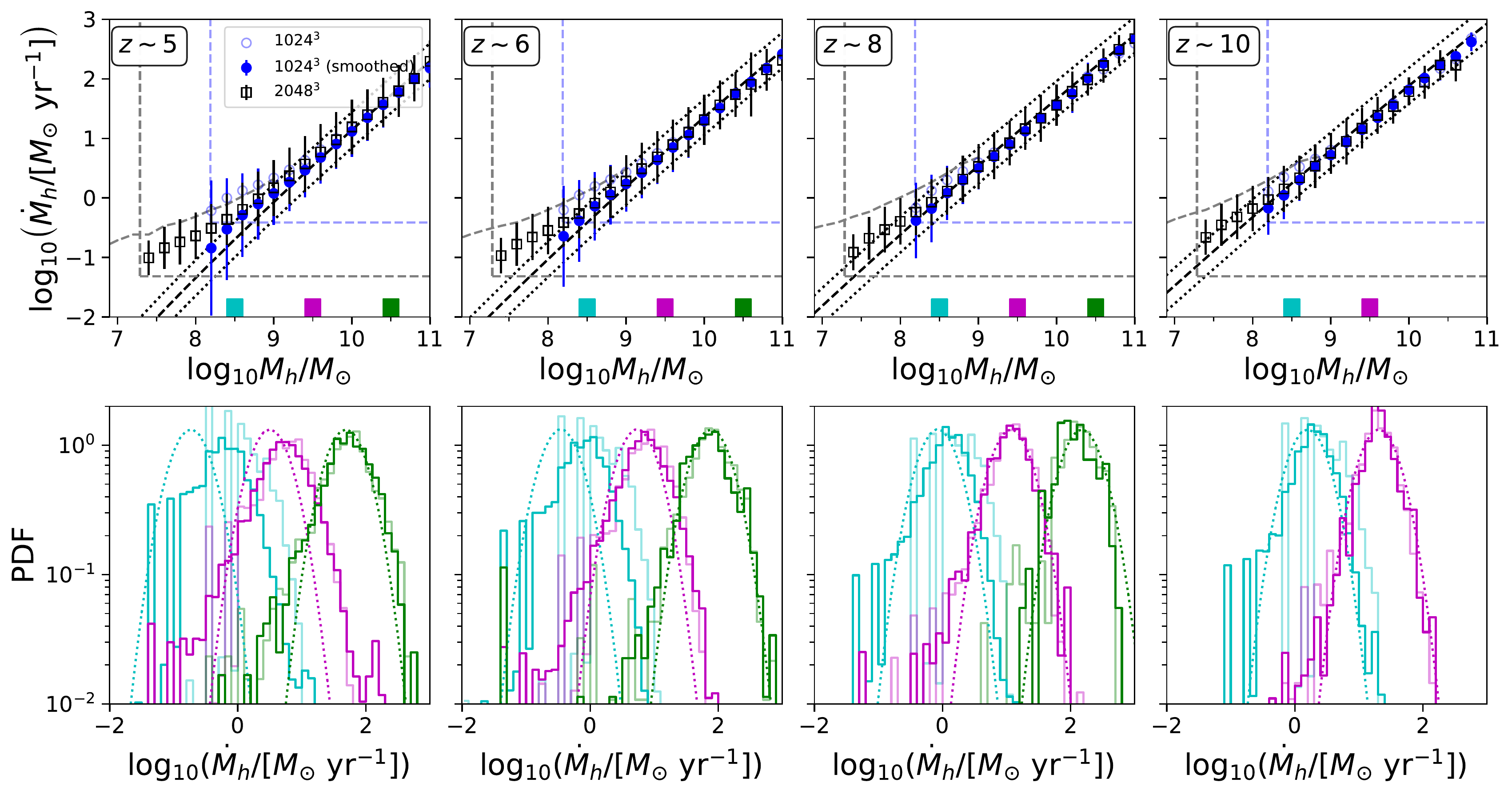}
\caption{{\bf Impact of low resolution on the mass accretion rates of low-mass halos.} Structure is identical to that of Figure \ref{fig:hmar}: top panels show the MAR averaged in a series of halo mass bins, as computed via $1024^3$ (blue) and $2048^3$ (black) particle $N$-body simulations in a $(40 \ h^{-1} \mathrm{Mpc})^3$ volume, while bottom panels show the full distribution of the MAR in three 0.2 dex mass bins centered on $\log_{10} M_h = 8.5, 9.5$, and 10.5 (cyan, magenta, green; indicated along $x$-axis in top row). Resolution limits are indicated via dashed lines in the top row, including the halo mass resolution (20 particles; vertical line), accretion rate resolution (1 particle per timestep; horizontal line), as well as an estimate of shot noise in the accretion rate (see \S\ref{sec:rhs} for details). Smoothing of the MAR over 100 Myr timescales (filled blue; top row) results in MAR-$M_h$ relations more in-line with the high resolution simulation results (black), though these results still depart from the \citet{Furlanetto2017} analytic model at $M_h \lesssim 10^9 \ M_{\odot}$.}
\label{fig:marconvergence}
\end{center}
\end{figure*}

Figure A1 shows the results of this convergence study. The structure is identical to Figure \ref{fig:hmar}: the top row shows the relationship between halo mass and MAR, while the bottom row shows the full PDF of the MAR in select mass bins. We investigate the effects of resolution and smoothing. Starting in the top row, we see once again the limitations of the low-resolution simulation at $z \lesssim 8$, with MARs asymptoting toward the MAR resolution limit of one particle per timestep (semi-transparent blue points). Black points show the results obtained from the higher resolution simulation; a departure from a pure power-law $\dot{M}_h$--$M_h$ relation is apparent at $M_h \lesssim 10^9 \ \Msun$, even for the high resolution simulation. This could be an important consideration for models investigating signatures of, e.g., Pop~III stars in minihalos before reionization.

\end{document}